\documentclass[draftclsnofoot,onecolumn, 12pt]{IEEEtran}
\usepackage{amsmath,amssymb,amsfonts}
\usepackage[ruled,vlined,linesnumbered]{algorithm2e}
\usepackage{algpseudocode}
\usepackage{xr}
\usepackage{pifont}
\usepackage{tabularx}
\usepackage{cellspace}
\makeatletter
\newcommand{\removelatexerror}
{\let\@latex@error\@gobble}
\makeatother
\usepackage{graphicx}
\usepackage{textcomp}
\usepackage{xcolor}
\usepackage{array}
\usepackage{textcomp}
\usepackage{dutchcal}
\usepackage{multirow}
\usepackage{bm}
\usepackage{booktabs}
\usepackage{ntheorem}
\usepackage{subfigure} 
\usepackage{orcidlink} 
\hypersetup{hidelinks}
\usepackage[mathscr]{euscript}
\usepackage{euscript}
\usepackage{mathtools}
\usepackage{epstopdf}
\usepackage{cases}
\def\BibTeX{{\rm B\kern-.05em{\sc i\kern-.025em b}\kern-.08em
		T\kern-.1667em\lower.7ex\hbox{E}\kern-.125emX}}
\usepackage{cases}
\theoremseparator{.}

\theorembodyfont{}

\begin{document}
	
\title{Generative AI-enhanced Low-Altitude UAV-Mounted Stacked Intelligent Metasurfaces}

\author{
Geng~Sun$^{\orcidlink{0000-0001-7802-4908}}$,~\IEEEmembership{Senior Member,~IEEE,}
Mingzhe~Fan,\IEEEmembership{}
Lei~Zhang, 
Hongyang~Pan$^{\orcidlink{0000-0003-4948-8062}}$,\IEEEmembership{}
Jiahui~Li$^{\orcidlink{0000-0002-7454-3257}}$,\IEEEmembership{}
Chuang~Zhang$^{\orcidlink{0000-0002-0505-0512}}$,\IEEEmembership{}
Linyao~Li,\IEEEmembership{}
Changyuan~Zhao$^{\orcidlink{0000-0001-9187-9572}}$,\IEEEmembership{}
and Chau~Yuen$^{\orcidlink{0000-0002-9307-2120}}$,~\IEEEmembership{Fellow,~IEEE}
\thanks{This study is supported in part by the National Natural Science Foundation of China (62272194, 62471200), in part by the Science and Technology Development Plan Project of Jilin Province (20250101027JJ), in part by the Postdoctoral Fellowship Program of China Postdoctoral Science Foundation (GZC20240592), in part by China Postdoctoral Science Foundation General Fund (2024M761123), and in part by the Scientific Research Project of Jilin Provincial Department of Education (JJKH20250117KJ). (\textit{Corresponding author: Jiahui Li.)}}
\thanks{Geng Sun is with the College of Computer Science and Technology, Key Laboratory of Symbolic Computation and Knowledge Engineering of Ministry of Education, Jilin University, Changchun 130012, China, and also with the College of Computing and Data Science, Nanyang Technological University, Singapore 639798 (e-mail: sungeng@jlu.edu.cn).}

\thanks{Mingzhe Fan and Lei Zhang are with the College of Computer Science and Technology, Jilin University, Changchun 130012, China (e-mails: fanmz24@mails.jlu.edu.cn, zhanglei\_2123@163.com).}

\thanks{Hongyang Pan is with the College of Information
Science and Technology, Dalian Maritime University, Dalian 116026,
China (e-mail: panhongyang18@foxmail.com).}

\thanks{Jiahui Li is with the College of Computer Science and Technology, and also with the Key Laboratory of Symbolic Computation and Knowledge Engineering of Ministry of Education, Jilin University, Changchun 130012, China (e-mail: lijiahui@jlu.edu.cn).}

\thanks{Chuang Zhang is with the College of Computer Science and Technology, Jilin University, Changchun 130012, China, and also with the Singapore University of Technology and Design, Singapore 487372 (email: chuangzhang1999@gmail.com).}

\thanks{Linyao Li is with the College of Software Engineering, Jilin University, Changchun 130012, China (e-mail: lily2421@mails.jlu.edu.cn).}
\thanks{Changyuan Zhao is with the College of Computing and Data Science, Nanyang Technological University, Singapore, and CNRS@CREATE, 1 Create Way, 08-01 Create Tower, Singapore 138602 (e-mail: zhao0441@e.ntu.edu.sg).}
\thanks{Chau Yuen is with the School of Electrical and Electronics Engineering, Nanyang Technological University, Singapore 639798 (e-mail: chau.yuen@ntu.edu.sg).}

}

\markboth{Journal of \LaTeX\ Class Files,~Vol.~X, No.~X, APRIL~2025}%
{Shell \MakeLowercase{\textit{et al.}}: Bare Demo of IEEEtran.cls for Computer Society Journals}
	
\IEEEtitleabstractindextext{%
\begin{abstract}		
Wireless communication systems face challenges in meeting the demand for higher data rates and reliable connectivity in complex environments. Stacked intelligent metasurfaces (SIMs) have emerged as a promising technology for advanced wave-domain signal processing, where mobile SIMs can outperform fixed counterparts. In this paper, we propose a novel unmanned aerial vehicle (UAV)-mounted SIM (UAV-SIM) assisted communication system within low-altitude economy (LAE) networks, where UAVs act as both cache-enabled base stations and mobile SIM carriers to enhance uplink transmissions. To maximize network capacity, we formulate a UAV-SIM-based joint optimization problem (USBJOP) that integrates user association, UAV-SIM three-dimensional positioning, and multi-layer SIM phase shift design. Due to the non-convexity and NP-hardness of USBJOP, we decompose it into three subproblems, which are the association between UAV-SIMs and users optimization problem (AUUOP), the UAV location optimization problem (ULOP), and the UAV-SIM phase shifts optimization problem (USPSOP). Then, we solve them through an alternating optimization strategy. Specifically, AUUOP and ULOP are transformed into convex forms solvable via the CVX tool, while USPSOP is addressed by a generative artificial intelligence (GAI)-based hybrid optimization algorithm. Simulation results show that the proposed approach achieves approximately 1.5 times higher network capacity compared with suboptimal schemes, effectively mitigates multi-user interference with increasing SIM layers and meta-atoms, and reduces runtime by 10\% while maintaining solution quality, thereby demonstrating its practicality for real-world deployments.
\end{abstract}

}
\maketitle
\IEEEdisplaynontitleabstractindextext
\IEEEpeerreviewmaketitle
%
\section{Introduction}
\label{sec:introduction}

\par \IEEEPARstart{W}ith the rapid evolution of wireless communication technologies, traditional communication architectures continue to face significant challenges in terms of flexibility, coverage, and computational efficiency~\cite{Pan2025}. Although fifth-generation (5G) and beyond networks, such as low-altitude economy (LAE) networks, have made substantial progress in addressing these limitations, they still struggle with providing reliable service in dynamic environments and meeting the growing demands for real-time signal processing capabilities~\cite{Soderi2024}. In this case, reconfigurable intelligent surfaces (RISs) have emerged as a promising solution for enhancing wireless communication by manipulating electromagnetic (EM) waves through programmable meta-elements~\cite{Wang2025}. However, conventional single-layer RIS structures face inherent limitations in performing complex signal processing tasks, which has led to the development of more advanced architectures such as stacked intelligent metasurfaces (SIMs)~\cite{An2023a}.

\par Specifically, SIMs represent a significant advancement over traditional RIS technology by adopting a multi-layer architecture that mimics the structure of artificial neural networks~\cite{An2024}. Different from conventional single-layer RIS that merely reflects incident signals with adjustable phase shifts, the multi-layer SIM architecture employs multiple cascaded metasurface layers that exponentially increase the available degrees of freedom for electromagnetic wave manipulation. This multi-layer configuration enables SIMs to perform sophisticated EM wave manipulations and computational tasks at the speed of light through direct wave-domain computation, thereby eliminating the processing delays typically associated with conventional digital signal processing methods while achieving significantly enhanced capabilities compared to single-layer structures~\cite{An2023}. This capability is particularly crucial in mobile platform deployment scenarios where real-time adaptation to changing channel conditions is essential, as traditional single-layer RIS requires frequent beamforming optimization recalculations when platform positions change. Additionally, SIMs can achieve comparable beamforming performance while requiring fewer radio frequency chains, thus reducing the system dependency on high-precision analog-to-digital and digital-to-analog conversion components~\cite{An2023a}. Furthermore, this enhanced control enables much more precise manipulation of wavefronts and effective multi-user interference suppression compared to the limited capabilities of single-layer designs. These characteristics make SIMs particularly attractive for applications requiring ultra-low latency and high computational efficiency in wireless networks.

\par Despite their promising capabilities, SIM implementations face significant practical challenges when deployed with fixed base stations (BSs). Since SIM systems generally lack cache modules, they perform optimally when integrated with dedicated BSs~\cite{Dai2025,Dai2025a}. However, fixed BS deployments inherently limit the potential of SIM technology by constraining coverage area, adaptability to changing network conditions, and ability to serve mobile users efficiently. These limitations are particularly pronounced in scenarios requiring rapid deployment, such as emergency response situations, temporary event coverage, or service provision in remote areas where fixed infrastructure is impractical or cost-prohibitive~\cite{Li2024a}. Furthermore, fixed deployments cannot easily adapt to dynamic user distributions and changing network traffic patterns, resulting in suboptimal resource utilization and reduced overall network performance~\cite{Sun2025}.

\par To overcome these limitations, mounting SIMs on mobile platforms within the LAE networks presents a compelling solution that could significantly enhance their versatility and performance. Among various mobile platform options, unmanned aerial vehicles (UAVs) have garnered particular interest due to their exceptional mobility, low deployment cost, and ability to establish favorable line-of-sight (LoS) communication channels with ground users~\cite{Li2024c}. In particular, several application domains can significantly benefit from UAV-assisted mobile communication platforms. For example, in emergency response operations such as earthquakes or floods where terrestrial communication infrastructure is often severely damaged, UAV-based systems can be rapidly deployed to reconfigure coverage and maintain connectivity for geographically dispersed survivors~\cite{Zheng2025}. In the agricultural sector, UAV-assisted platforms can flexibly patrol farmland and dynamically extend coverage to temporary blind spots or high-demand areas, thereby supporting high-resolution sensing and imagery transmission~\cite{Wang2025d}. Similarly, in urban environments characterized by dense user distributions and severe signal blockage from high-rise buildings, UAV-based mobile platforms can adaptively adjust both positions and communication parameters to enhance traffic management and public safety services~\cite{Cheng2025}. As such, the superior LoS conditions facilitated by UAVs in the LAE networks are especially beneficial for SIM-based wave-domain calculations, as they reduce the complexity of channel estimation and improve the accuracy of signal processing operations. Moreover, UAVs possess a certain cache capacity, enabling them to store data processed through SIM-based computations, thus complementing the lack of storage capabilities of SIMs~\cite{Zhang2025}. This synergistic integration creates the UAV-mounted SIMs (UAV-SIMs), which represent a promising technology for next-generation wireless networks.

\par However, achieving such potential of UAV-SIM systems in LAE networks presents several significant challenges that require innovative solutions. \textit{Firstly}, the joint optimization of user-UAV associations and UAV positions introduces a complex decision space with numerous interdependencies, as changes in UAV locations affect the quality of service for multiple users simultaneously~\cite{Wu2018}. \textit{Secondly}, optimizing the phase shifts of multiple SIM layers adds another dimension of complexity, which requires sophisticated algorithms that can effectively configure thousands of meta-elements across multiple stacked layers~\cite{Lin2024}. \textit{Finally}, the integration of these optimization problems creates a highly non-convex solution space with multiple local optima, making conventional optimization approaches ineffective for real-world deployments. These challenges are further compounded by the energy constraints of UAVs, which necessitate efficient resource allocation strategies to prolong operational time while maintaining high-quality communication links.

\par Thus, in this paper, we propose a comprehensive framework for UAV-SIM-assisted uplink communications that addresses these challenges through a novel alternating optimization method with generative artificial intelligence (GAI)~\cite{Wang2024b}. Our key contributions are as follows:

\begin{itemize}
    \item \textit{UAV-SIM Architecture for Uplink Communications:} We design a novel system architecture that leverages multiple UAV-SIMs to enhance uplink communication performance for ground users. This architecture uniquely integrates the computational capabilities of multi-layer SIMs with the mobility advantages of UAVs, creating a flexible and high-performance communication platform. To the best of our knowledge, such a platform can significantly reduce latency during natural disaster monitoring, thereby enabling low-cost and low-power signal transmission, which extends the service duration of the system~\cite{Liu2025b}.

    \item \textit{Multi-dimensional Optimization Problem Formulation:} We formulate a UAV-SIM-based joint optimization problem (USBJOP) that simultaneously considers three critical dimensions, which are the association between UAV-SIMs and ground users, the three-dimensional positioning of UAV-SIMs, and the phase shift configurations across multiple SIM layers. This formulation captures the complex interplay between these variables and their collective impact on network capacity. Moreover, USBJOP is proven to be NP-hard and non-convex.

    \item \textit{Alternating Optimization Method with GAI:} To tackle the inherent complexity of the USBJOP, which is both non-convex and NP-hard, we propose a novel alternating optimization (AO) strategy that decomposes the original problem into three manageable sub-problems, \textit{i.e.}, the association between UAV-SIMs and users optimization problem (AUUOP), the UAV location optimization problem (ULOP), and the UAV-SIM phase shifts optimization problem (USPSOP). Following this, we solve the AUUOP and ULOP via convex optimization, and address the USPSOP with a GAI-based hybrid optimization algorithm~\cite{Li2025a}. Such a GAI method can accelerate the solution of its sub-problem within the AO framework, thus reducing the algorithm runtime in complex scenarios while ensuring high solution quality.

    \item \textit{Performance Analysis:} Simulation results demonstrate that UAV-SIMs utilizing the proposed AO strategy achieve approximately $1.5$ times the network capacity compared to suboptimal benchmarks. Moreover, the GAI method increases the overall speed of the AO algorithm by $10\%$ in complex scenarios, while ensuring the quality of the final solution. We further analyze the effectiveness of our designed loss function with the capacity-oriented term, which maintains consistent performance advantages ranging from $20\%$ to $400\%$ across all configurations. In addition, we investigate the impact of key system parameters, including the number of SIM layers, the number of meta-atoms on each layer, and user density, revealing that the SIM is highly effective at mitigating multi-user interference, and this effectiveness is positively correlated with the number of SIM layers and meta-atoms per layer within a specific range. Finally, we expand the simulation scenarios and find that the system continues to exhibit these favorable characteristics.
\end{itemize}

\par The remainder of this paper is organized as follows. Section \ref{sec3} introduces the system model and formulates the USBJOP. In Section \ref{sec4}, we present our alternating optimization strategy to solve the formulated problem. Section \ref{sec5} provides comprehensive simulation results and performance analysis. Finally, we conclude the paper and discuss future research directions in Section \ref{sec6}.

\par \textit{Notations:} In this paper, $(\cdot)^{\mathrm{H}}$ and $(\cdot)^{\mathrm{T}}$ denote the Hermitian transpose and transpose, respectively; bold uppercase and lowercase letters denote matrices and vectors, respectively; $ \operatorname{diag}(\cdot)$ means a diagonal matrix; the smallest integer greater than or equal to $x$ is denoted as $\lceil x\rceil$; the space of $x \times y$ complex-valued matrices is represented by $\mathbb{C}^{x \times y}$; $\|\mathbf{x}\|$ represents the norm of the vector $x$; mod $(x, n)$ returns the remainder after division of $x$ by $n$; the distribution of a circularly symmetric complex Gaussian random vector with mean vector $\boldsymbol{\mu}$ and covariance matrix $\boldsymbol{\Xi} \succeq \mathbf{0}$ is denoted by $\sim \mathscr{CN}(\boldsymbol{\mu}, \boldsymbol{\Xi})$, where $\sim$ stands for ``distributed as''; $\operatorname{sinc}(x)=\frac{\sin (\pi x)}{\pi x}$ is the normalized sinc function; $j$ is the imaginary unit satisfying $j^2=-1$.

%
\section{Related Works}

\par In this section, we review the existing literature related to this paper on UAV-SIM-assisted uplink communications within LAE networks. Specifically, the literature review is organized into three main categories, including SIM architectures, UAV-assisted wireless communications, and optimization methods in SIM and UAV deployments. To further highlight the differences between this paper and previous works, we summarize a comparison table as shown in Table \ref{RWComparison}.

%
\subsection{SIM Architectures}

\par Intelligent metasurfaces have emerged as a transformative technology for wireless communications due to their ability to manipulate electromagnetic waves through programmable meta-elements. The authors in~\cite{Mu2021} introduced the fundamental principles of RISs, which highlight their potential to enhance signal quality through passive beamforming while maintaining low power consumption. Moreover, the authors in~\cite{Wu2020} demonstrated that RIS technology can significantly improve spectral and energy efficiency by creating favorable propagation environments without requiring additional active components. However, these conventional single-layer metasurface designs face inherent limitations in performing complex signal processing tasks due to their limited degrees of freedom and passive nature.

\par \par To overcome these limitations, SIMs have been proposed as an advanced architecture with enhanced capabilities. The authors in~\cite{An2023a} introduced the concept of SIMs with multiple cascaded layers that mimic the structure of neural networks, enabling sophisticated wave-domain computing capabilities at the speed of light. Furthermore, the authors in~\cite{Papazafeiropoulos2024a} provided a thorough analytical characterization of multi-layer metasurface structures, demonstrating their superior beamforming precision and enhanced signal processing capabilities compared to single-layer designs. Additionally, the authors in~\cite{Li2024} developed novel optimization techniques specifically designed for configuring the phase shifts across multiple SIM layers, achieving significant performance improvements for complex electromagnetic wave manipulation tasks. Besides, the authors in~\cite{An2024a} utilized SIM to perform optical two-dimensional discrete Fourier transform for direction-of-arrival estimation, enabling direct angular spectrum observation without radio frequency chains. Nevertheless, most existing research on SIMs focuses on fixed deployments with dedicated base stations, which significantly limits their adaptability to dynamic communication environments and changing user distributions within the context of LAE networks.

%
\subsection{UAV-assisted Wireless Communications}

\par In the LAE networks, UAVs have gained significant attention in wireless communications due to their mobility, deployment flexibility, and ability to establish favorable LoS channels with ground users. Specifically, the authors in~\cite{Nguyen2018} proposed to employ cooperative non-orthogonal multiple access and jointly optimize radio resources with the positions of the UAVs, thereby maximizing the sum rate of all users. Moreover, the authors in~\cite{Yu2021} applied deep reinforcement learning for the multi-objective optimization of the data rate, the harvested energy, and the energy consumption of the UAV. In addition, the authors in~\cite{Xu2021} investigated a UAV-assisted multi-access edge computing system and jointly optimized the task assignment, resource allocation, and the trajectory of the UAV, thereby maximizing the weighted computation efficiency of the system. Besides, the author in~\cite{Ali2020} studied a UAV-mounted aerial BS and optimized its 3D placement and power allocation to enhance the minimum data rates of both aerial and terrestrial users, and demonstrated the performance gains of their approach over several benchmark schemes. However, these studies primarily employed UAVs equipped with conventional communication hardware, which limits their signal processing capabilities and spectral efficiency.

\par Several research efforts have specifically focused on optimizing the positioning and resource allocation of UAV-based communication systems. The authors in~\cite{Wu2018} aimed to maximize the minimum throughput of users by jointly optimizing user scheduling and association, the trajectory of UAVs, and power control, for which an iterative algorithm based on block coordinate descent and successive convex optimization was proposed. Furthermore, the authors in~\cite{Luong2021} proposed a novel algorithm where the positions of UAVs are determined using a deep Q-learning approach, a method specifically designed to find optimal locations without requiring channel state information (CSI). In addition, the authors in~\cite{Kim2024} investigated a general framework for the deployment of UAVs. Their proposed method optimizes the locations of UAVs for any given performance metric by using a DNN-based surrogate model combined with zeroth-order optimization, which allows the predicted performance of the network to guide the physical placement of the UAVs. Moreover, the authors in~\cite{Sun2024} designed a three-layer post-disaster rescue computing architecture integrating mobile edge computing and vehicle fog computing, and proposed a joint task offloading and resource allocation algorithm, which outperformed benchmark methods under heavy workloads. Nevertheless, these approaches generally consider UAVs with standard antenna configurations rather than advanced signal processing technologies like intelligent metasurfaces.

%
\subsection{Optimization Methods in SIM and UAV Deployments}

\par Optimizing the configuration and deployment of both SIMs and UAVs presents significant challenges due to their complex operational characteristics and interdependencies. For SIM optimization, the authors in~\cite{Shi2025} proposed a wave-based beamforming algorithm for SIMs in cell-free massive multiple-input multiple-output (mMIMO) systems that relies only on statistical CSI, which can significantly enhance the spectral efficiency. Moreover, the authors in~\cite{Papazafeiropoulos2025a} developed a mMIMO setup with SIMs at both the BS and in the intermediate space and proposed an approach to simultaneously optimize the phase shifts of both SIMs. In addition, the authors in~\cite{Liu2025} presented an optimization formulation for
the joint design of the SIM phase shifts and the transmit power allocation, which is efficiently tackled via a customized deep reinforcement learning approach. However, these works focused exclusively on fixed SIM, neglecting the additional complexity introduced by mobile platforms such as UAVs.

\par For UAV deployment optimization, the authors in~\cite{Qiu2020} considered the deployment of multiple UAV-mounted BS and addressed the joint problem of placement, resource allocation, and user association with constrained backhaul links, for which an efficient iterative algorithm is developed to optimize user association and UAV placement jointly. Furthermore, the authors in~\cite{Zhang2021} developed a three-step method to minimize the number of required UAVs and improve the rate of coverage by optimizing the three-dimensional (3D) positions of the UAVs, the clustering of users, and the allocation of the frequency band. Nevertheless, these studies primarily considered standard communication equipment on UAVs rather than advanced technologies like SIMs, which introduce additional optimization variables and constraints.

\par A limited number of works have begun exploring the integration of intelligent surfaces with mobile aerial platforms in LAE networks. The authors in~\cite{Tyrovolas2022} investigated the utilization of a UAV-mounted RIS for data collection in internet-of-things networks, studying the coverage probability and proposing an energy model to analyze the performance of the data collection scheme. Additionally, the authors in~\cite{Li2024b} provided a novel UAV-RIS-aided interference management scheme for space-air-ground integrated networks to mitigate diverse interference, where the UAV-RIS is introduced for the cooperating interference elimination process, which can improve the capacity of the system. Despite these advances, existing research lacks comprehensive optimization frameworks that simultaneously address user association, platform positioning, and metasurface configuration for UAV-SIM systems. Most importantly, no prior work has thoroughly investigated the joint optimization of these interdependent variables while considering the unique characteristics of stacked metasurfaces in mobile aerial deployments. Different from these works, we aim to investigate the UAV-SIM-assisted uplink communication system in LAE networks and propose a efficient alternating optimization strategy that makes this challenging problem tractable. 


\begin{table*}[htb]
\renewcommand{\arraystretch}{0.9}
    \centering
    \caption{Comparison between related works and our contribution}
    \label{RWComparison}
    \begin{tabular*}{\textwidth}{@{}@{\extracolsep{\fill}}cccccccc@{}}
\toprule
\textbf{Reference} & \begin{tabular}[c]{@{}c@{}}\textbf{UAV}\end{tabular} & \begin{tabular}[c]{@{}c@{}}\textbf{SIM/RIS}\end{tabular} & \begin{tabular}[c]{@{}c@{}}\textbf{Optimization} \\ \textbf{Method}\end{tabular} & \begin{tabular}[c]{@{}c@{}}\textbf{Communication} \\ \textbf{Metrics}\end{tabular} & \begin{tabular}[c]{@{}c@{}}\textbf{User}\\ \textbf{Association}\end{tabular} & \begin{tabular}[c]{@{}c@{}}\textbf{3D}\\ \textbf{Positioning}\end{tabular} & \begin{tabular}[c]{@{}c@{}}\textbf{SIM/RIS}\\ \textbf{Phase Shifts}\end{tabular} \\ \hline \hline
\cite{Mu2021}  & $\times$  & $\checkmark$  & Convex Optimization  & Network Capacity & $\checkmark$  & $\times$ & $\checkmark$  \\
\cite{Wu2020}  & $\times$  & $\checkmark$  & Convex Optimization  & Transmit Power & $\checkmark$  & $\times$ & $\checkmark$  \\
\cite{An2023a}  & $\times$  & $\checkmark$  & Traditional  & Error Minimization & $\times$  & $\times$ & $\checkmark$  \\
\cite{An2024a}  & $\times$  & $\checkmark$  & Traditional  & DOA Estimation & $\times$  & $\times$ & $\checkmark$  \\
\cite{Nguyen2018}  & $\checkmark$  & $\times$  & 	Convex Optimization  & Sum Rate & $\checkmark$  & $\checkmark$ & $\times$  \\
\cite{Yu2021}  & $\checkmark$  & $\times$  & Learning-based  & Energy Efficiency & $\checkmark$  & $\checkmark$ & $\times$  \\
\cite{Wu2018}  & $\checkmark$  & $\times$  & Convex Optimization  & Min Throughput & $\checkmark$  & $\checkmark$ & $\times$  \\ 
\cite{Sun2024}  & $\checkmark$  & $\times$  & Heuristic  & Task Completion & $\checkmark$  & $\times$ & $\times$  \\
\cite{Shi2025}  & $\times$  & $\checkmark$  & Traditional  & Spectral Efficiency & $\checkmark$  & $\times$ & $\checkmark$  \\
\cite{Qiu2020}  & $\checkmark$  & $\times$  & Traditional  & User Throughput & $\checkmark$  & $\checkmark$ & $\times$  \\
\cite{Tyrovolas2022}  & $\checkmark$  & $\checkmark$  & Traditional  & Coverage Probability & $\checkmark$  & $\times$ & $\times$  \\
\cite{Li2024b}  & $\checkmark$  & $\checkmark$  & Traditional  & System Capacity & $\times$  & $\times$ & $\times$  \\
\textbf{This Work}  & $\checkmark$  & $\checkmark$  & Convex Optimization  & Sum Rate & $\checkmark$  & $\checkmark$ & $\checkmark$  \\
\bottomrule
\end{tabular*}
\end{table*}

\begin{figure*}[!t]
    \centering
    \includegraphics[width=\textwidth]{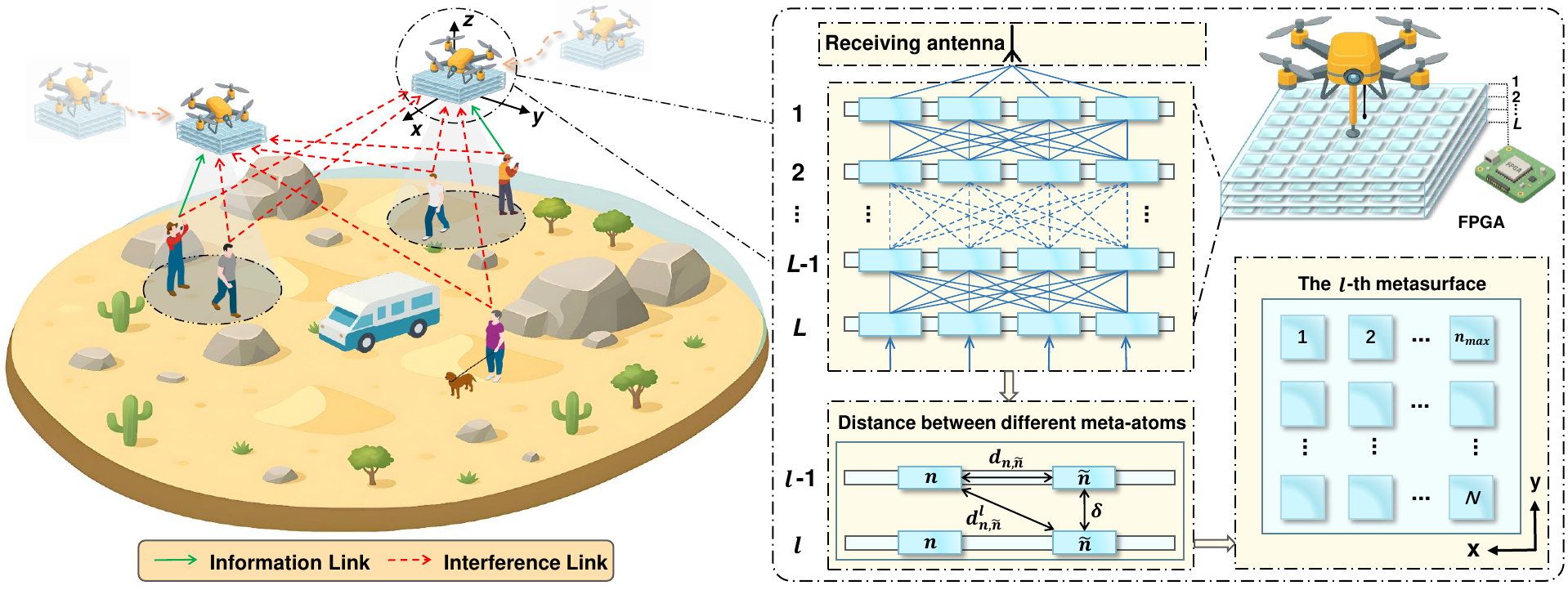}
    \caption{An uplink communication system assisted by UAV-SIM within the LAE networks paradigm, in which multiple UAV-SIMs are deployed at a fixed altitude while maintaining a safe distance, to process data from multiple ground users.}
    \label{System_model} 
    \vspace{-5pt}
\end{figure*}

%
\section{Models and Problem Formulation}\label{sec3}
\par In this section, we first present an overview of the considered UAV-SIM-assisted uplink communication system within LAE networks. Then, we detail the SIM transmission model including channel characteristics and signal reception. Finally, we introduce the UAV energy consumption model that quantifies the operational costs of our proposed aerial platform.

%
\subsection{System Overview}

\par As depicted in Fig. \ref{System_model}, we consider an uplink communication scenario assisted by UAV-SIMs within the paradigm of LAE networks. Specifically, within a finite area, a group of UAV-SIMs $\mathcal{M} = \{1, \ldots, M\}$ are deployed at a fixed altitude to serve a set of ground users $\mathcal{K} = \{1, \ldots, K\}$. Without loss of generality, we consider that $M<K$, which is common in practical applications where a limited number of UAVs are deployed to serve a larger user base. Given the physical constraints of actual deployment, such as limited payload, size, and power budget, each UAV-SIM is equipped with a single antenna. Meanwhile, to maximize spectrum utilization, particularly in scenarios with scarce spectrum, all UAV-SIMs operate on a shared frequency, which leads to mutual interference when multiple users communicate simultaneously. Additionally, the specific physical structure of the SIM is presented in Fig. \ref{System_model}, showing the stacked arrangement of multiple metasurface layers.

\par Adopting the three-dimension (3D) Cartesian coordinate system in the context of LAE networks, we consider that the $k$-th user is fixed at $\boldsymbol{u}_k=[x_k,y_k,0]^{\mathrm{T}}, k \in \mathcal{K}$. Since all UAV-SIMs fly at a fixed altitude $H$ throughout the LAE networking environment, the coordinate of the $m$-th UAV-SIM can be represented as $\boldsymbol{w}_m=[x_m,y_m,H]^{\mathrm{T}}, m \in \mathcal{M}$. Furthermore, to mitigate the risk of collisions among UAV-SIMs and to prevent signal interference resulting from insufficient separation, it is necessary to uphold a safe distance, which can be expressed as follows:
\begin{equation}
    \mathcal{C}_1 : \left\|\boldsymbol{w}_i-\boldsymbol{w}_j\right\| \geq D_{\min }, i \neq j, \forall i \in \mathcal{M}, \forall j \in \mathcal{M},
\end{equation}
\noindent where $D_{\min}$ represents the minimum safety distance between UAV-SIMs. Likewise, the distance from the $m$-th UAV-SIM to the $k$-th user is given by $D_{m, k}=\left\|\boldsymbol{w}_m-\boldsymbol{u}_k\right\|$.

\par Similar to \cite{Wu2018}, we consider that a UAV-SIM can serve one user and a user can only be served by one UAV-SIM. Note that this one-to-one association model is adopted to establish a foundational framework for the UAV-SIM assisted communication system, which allows us to clearly delineate the core optimization challenges without the added complexity of multi-user resource allocation mechanisms and enables precise analysis of the impact of various system parameters on network capacity. Furthermore, this model can be readily extended to accommodate multi-user scenarios through time-division multiple access~\cite{Wu2018} schemes where a single UAV-SIM serves multiple users in different time slots by reconfiguring its phase shifts accordingly. As such, for the considered model, mathematically we define a binary variable $S_{m, k}$, where $S_{m, k}=1$ means that there is an association between the $m$-th UAV-SIM and $k$-th user within LAE networks, \textit{i.e.}, $k$-th user can transmit data to the $m$-th UAV-SIM; otherwise, $S_{m, k}=0$, meaning that the $m$-th UAV-SIM does not provide service to the $k$-th user. The corresponding constraints can be written as follows:
\begin{equation}
    \mathcal{C}_2: \sum\nolimits_{m=1}^M S_{m, k} \leq 1, \forall k\in \mathcal{K},
\end{equation}
\begin{equation}
    \mathcal{C}_3: \sum\nolimits_{k=1}^K S_{m, k} \leq 1, \forall m\in \mathcal{M},
\end{equation}
\begin{equation}
    \mathcal{C}_4: S_{m, k} \in \{0,1\}, \forall m\in \mathcal{M}, \forall k\in \mathcal{K}.
\end{equation}

%
\subsection{SIM Transmission Model}

\par Let $\mathcal{L}=\{1, 2,..., L\}$ represent the set of metasurface layers, and $\mathcal{N}=\{1, 2,..., N\}$ represent the set of meta-atoms. Generally, we consider $\phi_n^l=\mathrm{e}^{j \theta_n^l}$ denotes the transmission coefficient imposed by the $n$-th meta-atom on the $l$-th metasurface layer, where $\theta_n^l$ denotes the corresponding phase shift, which satisfies $\theta_n^l \in[0,2 \pi), \forall n \in \mathcal{N}, \forall l \in \mathcal{L}$. Consequently, the diagonal phase shift matrix $\mathbf{\Phi}^l$ of the $l$-th metasurface layer is given as follows~\cite{Xie2024}:
\begin{equation}	\mathbf{\Phi}^l=\operatorname{diag}\left(\mathrm{e}^{j \theta_1^l}, \mathrm{e}^{j \theta_2^l}, \cdots, \mathrm{e}^{j \theta_N^l}\right) \in \mathbb{C}^{N \times N},\forall l \in \mathcal{L}.
\end{equation}

\par Without loss of generality, we consider that each metasurface is modeled as a uniformly planar array, and an isomorphic lattice structure is used to arrange all metasurface layers. As shown in Fig. \ref{System_model}, the element spacing between the $n$-th and $\tilde{n}$-th meta-atoms on the same metasurface layer can be written as follows \cite{An2023a}:
\begin{equation}
    d_{n, \tilde{n}}  =d_{e, n} \sqrt{\left(n_x-\tilde{n}_x\right)^2+\left(n_y-\tilde{n}_y\right)^2}, 
\end{equation}
\noindent where $d_{e, n}$ denotes the spacing between adjacent meta-atoms within the same metasurface layer. In addition, the indices of the $n$-th meta-atom in the $x$ and $y$ directions can be denoted as $n_x$ and $n_y$, respectively, which are defined as follows \cite{An2023a}:
\begin{equation}
    n_x=\bmod \left(n-1, n_{\max }\right)+1, n_y=\left\lceil n / n_{\max }\right\rceil,
\end{equation}
\noindent where $n_{\max}$ is the number of meta-atoms on each row of the metasurface. In this work, each metasurface is arranged in a square layout, and hence we have $N=n^2_{\max}$.

\par Let $\delta$ represent the distance between adjacent metasurface layers in the SIM, and $T_{\mathrm{SIM}}$ is the thickness of the SIM. Accordingly, we have $\delta=T_{\mathrm{SIM}}/L$, and the distance between any two meta-atoms located in different adjacent metasurface layers can be represented as follows~\cite{An2023a}:
\begin{equation}
    d_{n, \tilde{n}}^l=\sqrt{d_{n, \tilde{n}}^2+\delta^2}, \forall l \in \mathcal{L} /\{1\}.
\end{equation}

\par Next, we consider the transmission matrix between adjacent metasurfaces and from the transmitting antenna to the input metasurface of the SIM. Denote $\mathbf{W}^l \in \mathbb{C}^{N \times N},\forall l \in  \mathcal{L} /\{1\}$ as the channel response from the $(l-1)$-th metasurface layer to the $l$-th metasurface layer, and $\mathbf{w}^{1\mathrm{H}}_m \in \mathbb{C}^{1 \times N}$ is the channel response from the output metasurface layer of the SIM to the $m$-th UAV. According to the Rayleigh-Sommerfeld diffraction theory~\cite{Liu2022}, the response from the $\tilde{n}$-th meta-atom on the $(l-1)$-th metasurface layer to the $n$-th meta-atom on the $l$-th metasurface layer can be written as follows~\cite{An2023a}:
\begin{equation}
\begin{aligned}
w_{n, \tilde{n}}^l 
=\frac{d_{x}d_{y} \cos \psi_{n, \tilde{n}}^l}{d_{n, \tilde{n}}^l} 
  \left( \frac{1}{2 \pi d_{n, \tilde{n}}^l} - j \frac{1}{\lambda} \right) 
  \mathrm{e}^{j 2 \pi d_{n, \tilde{n}}^l / \lambda}, \quad \forall l \in \mathcal{L} \setminus \{1\},
\end{aligned}
\label{W}
\end{equation}

\noindent where $d_{n, \tilde{n}}^l$ is the transmission distance, $d_{x}d_{y}$ denotes the area of each meta-atom and $\psi_{n, \tilde{n}}^l$ is the angle between the propagation direction and the normal direction of the $(l-1)$-th metasurface layer. Similarly, the $n$-th entry $w^1_{n,m}$ of $\mathbf{w}^{1}_m$ can be obtained using Eq. (\ref{W}).

\par Therefore, the equivalent channel response at the $m$-th UAV-SIM can be written as follows \cite{Papazafeiropoulos2024a}:
\begin{equation}
    \mathbf{G}_{m}=\boldsymbol{\Phi}^L_{m} \mathbf{W}^L_{m} \boldsymbol{\Phi}^{L-1}_{m} \mathbf{W}^{L-1}_{m} \ldots \boldsymbol{\Phi}^2_{m} \mathbf{W}^2_{m} \boldsymbol{\Phi}^1_{m}  \in \mathbb{C}^{N \times N} .
\end{equation}

\par Let $\mathbf{h}_{m,k} \in \mathbb{C}^{N \times 1}$ denote the baseband equivalent channel from the $k$-th user to the input metasurface layer of the $m$-th UAV-SIM. Without loss of generality, a quasi-static flat-fading model is adopted for all channels, and hence $\mathbf{h}_{m,k}  \sim \mathscr{CN}\left(\mathbf{0}, \rho_0 D_{m, k}^{-2} \mathbf{R}\right), \forall m\in\mathcal{M}, \forall k\in\mathcal{K}\label{h}$ according to the correlated Rayleigh fading distribution, where $\rho_0$ denotes the channel power at the reference distance of $1$ m, and $\mathbf{R} \in \mathbb{C}^{N \times N}$ represents the spatial correlation matrix of the SIM. In an isotropic scattering environment, the spatial correlation matrix of the SIM for the far-field propagation is $\mathbf{R}_{n, \tilde{n}}=\operatorname{sinc}\left(2 d_{n, \tilde{n}} / \lambda\right), \forall \tilde{n} \in \mathcal{N},\forall n \in \mathcal{N},$ where $d_{n, \tilde{n}}$ represents the corresponding meta-atom spacing~\cite{An2023}.

\par  Then, the composite signal received by the $m$-th UAV-SIM can be written as follows~\cite{Wang2024a}:
{\begin{equation}
        r_{m,k} = \mathbf{w}_{m}^{1\mathrm{H}} \mathbf{G}_{m}^\mathrm{H} \sum\nolimits_{k=1}^K \mathbf{h}_{m,k}\sqrt{p_{k}} g_{k}+q_m,\forall m\in \mathcal{M}, \forall k\in \mathcal{K},
    \end{equation}
    \noindent where $p_k$ is the transmit power of the $k$-th user, $g_{k}$ is the signal from the $k$-th user and $q_m  \sim \mathscr{CN}\left(0, \sigma_m^2\right)$ denotes the additive white Gaussian noise (AWGN), as the multiuser interference is automatically mitigated as EM waves pass through the well-designed SIMs. Therefore, the signal-to-interference-plus-noise ratio (SINR) at the $m$-th UAV-SIM from the $k$-th user is given by \cite{An2023}:
    \begin{equation}	   \gamma_{m,k}=\frac{\left|\mathbf{w}_m^{1\mathrm{H}} \mathbf{G}_{m}^\mathrm{H} \mathbf{h}_{m,k}\right|^2 p_k }{\sum_{k^{\prime} \neq k}^K\left|\mathbf{w}_{m}^{1\mathrm{H}} \mathbf{G}_{m}^\mathrm{H} \mathbf{h}_{m,k^{\prime}}\right|^2 p_{k^{\prime}}+\sigma_m^2},\forall m\in \mathcal{M}, \forall k\in \mathcal{K},
    \end{equation}
    \noindent where $\sigma_m^2$ represents the average noise power of the $m$-th UAV. Then, the available rate at the $m$-th UAV-SIM from the $k$-th user is calculated as follows: 
    \begin{equation}
        R_{m,k} =  \log_2(1+\gamma_{m,k}).
    \end{equation}
  
    \par As a result, the network capacity can be expressed as:
    \begin{equation}	R=\sum\nolimits_{m=1}^M \sum\nolimits_{k=1}^K S_{m, k}R_{m,k}.
    \end{equation}

%
\subsection{UAV Energy Consumption Model}
\label{ssec:UAV energy}

\par The energy consumption of UAVs is a critical consideration in the design and operation of our UAV-SIM system, especially in LAE networks demanding energy-efficient and enduring UAV operations. For a rotary-wing UAV carrying a SIM payload, the total energy consumption consists primarily of propulsion energy and hovering energy. The propulsion energy consumption of the $m$-th UAV-SIM during flight can be modeled as follows \cite{Zeng2019}:
\begin{equation}
\begin{split}
P_m^{\text{prop}} = P_0 \left(1+\frac{3v_m^2}{U_{\text{tip}}^2}\right) + P_i\left(\sqrt{1+\frac{v_m^4}{4v_0^4}}-\frac{v_m^2}{2v_0^2}\right)^{1/2} + \frac{1}{2}d_0\rho s A v_m^3,
\end{split}
\end{equation}

\noindent where $P_0$ and $P_i$ represent the blade profile power and induced power in hovering status, respectively. The parameter $U_{\text{tip}}$ denotes the tip speed of the rotor blade, and $v_0$ refers to the mean rotor-induced velocity during hover. The UAV flying speed is represented by $v_m$, while $d_0$ is the fuselage drag ratio. Additionally, $\rho$ denotes the air density, $s$ is the rotor solidity, and $A$ represents the rotor disc area.

\par When a UAV-SIM is hovering at a fixed position to serve ground users, the energy consumption can be simplified to:
\begin{equation}
P_m^{\text{hover}} = P_0 + P_i,
\end{equation}

\noindent which represents the minimum power required to maintain the UAV altitude without horizontal movement.

\par Additionally, the SIM payload introduces an extra weight that affects the UAV energy efficiency. The total mass of the UAV-SIM system includes both the UAV mass and the mass of the SIM payload, which can be expressed as:
\begin{equation}
m_{\text{total}} = m_{\text{UAV}} + m_{\text{SIM}},
\end{equation}

\noindent where $m_{\text{UAV}}$ is the mass of the UAV platform, and $m_{\text{SIM}}$ is the mass of the SIM payload including all metasurface layers and supporting structures.

\par The added weight of the SIM payload increases the induced power $P_i$ required for hovering, which is proportional to the square root of the disk loading:
\begin{equation}
P_i = \kappa \sqrt{\frac{m_{\text{total}}^3}{A}},
\end{equation}

\noindent where $\kappa$ is a constant that depends on the air density and other aerodynamic parameters.

\par Given these energy consumption models, the relationship between the total energy consumption of UAV-SIM and its battery capacity can be expressed as:
\begin{equation}
\mathcal{C}_5:\left( P_m^{\text{hover}} + P_m^{\text{extra}} \right) T_{\text{operation}} + P^{\text{prop}}_{m}\frac{D_{\text{distance}}}{v_m} \leq E_{\text{battery}},
\end{equation}

}

\noindent where $E_{\text{battery}}$ is the UAV battery capacity, and $P_m^{\text{extra}}$ accounts for additional energy consumption from SIM configuration adjustments and communication module operations in LAE networks. The service time of UAV-SIM can be expressed as $T_{\text{operation}}$, and $D_{\text{distance}}$ indicates the flight distance of the UAV-SIM. This constraint highlights the importance of optimizing the UAV position to minimize repositioning needs while maximizing communication performance.

%
\subsection{Problem Formulation}

\par In this subsection, we formulate the optimization problem to maximize network capacity for our UAV-SIM-assisted uplink communication system within LAE networks. This requires jointly optimizing three key aspects as follows. \textit{First}, determining which users should be served by which UAV-SIMs. \textit{Second}, determining the optimal 3D positions of the UAV-SIMs. \textit{Third}, configuring the phase shifts across all metasurface layers to create favorable electromagnetic propagation conditions.
\par Specifically, by defining $\mathbf{S}\triangleq\left\{S_{m,k}, \forall m\in \mathcal{M}, \forall k\in \mathcal{K}\right\}$ as the association matrix between UAV-SIMs and ground users, $\mathbf{W}\triangleq \left\{\boldsymbol{w}_m, \forall m\in \mathcal{M} \right\}$ as the set of UAV-SIM positions, $\boldsymbol{\theta}^l \triangleq\left[\theta_1^l, \theta_2^l, \cdots, \theta_N^l\right]^{\mathrm{T}}$ as the phase shift configuration for layer $l$, and $\boldsymbol{\vartheta} \triangleq\left\{\boldsymbol{\theta}^1, \boldsymbol{\theta}^2, \cdots, \boldsymbol{\theta}^L\right\}$ as the complete phase shift configuration across all layers, the USBJOP can be formulated as follows:
\begin{subequations}\label{eq:2}
	\begin{align}
		\text {(USBJOP) }:\underset{\mathrm{\mathbf{S},\mathbf{W},\boldsymbol{\vartheta}}}{\max}\quad  &R  \\
		\text { s.t. } \quad
		& \mathcal{C}_1-\mathcal{C}_5,\\ 
		& \mathcal{C}_6:\theta_n^l \in[0,2 \pi), \forall n \in \mathcal{N}, \forall l \in \mathcal{L}. \label{phase shift constraint1}
	\end{align}
\end{subequations}

\noindent This optimization problem presents several significant challenges that make it difficult to solve directly. \textit{First}, it involves both discrete variables ($\mathbf{S}$) and continuous variables ($\mathbf{W}$ and $\boldsymbol{\vartheta}$), resulting in a non-linear mixed-integer programming problem. The discrete user association variables introduce combinatorial complexity, as there are $M^K$ possible associations to consider. \textit{Second}, the objective function is highly non-convex due to the complex relationship between UAV positions, phase shifts, and the resulting channel gains. This non-convexity arises from the intricate electromagnetic interactions within the multi-layer metasurface structure and the distance-dependent channel characteristics.

\par Furthermore, the constraints introduce additional complexity. Specifically, $\mathcal{C}_1$ ensures sufficient separation between UAVs to prevent collisions and interference in LAE networks, but creates non-convex feasible regions. Moreover,  $\mathcal{C}_2$-$\mathcal{C}_4$ enforce the one-to-one association between UAVs and users, introducing combinatorial challenges. In addition, $\mathcal{C}_6$ constrains the phase shift values within their physical limits. The combination of these constraints, particularly when considered alongside the energy constraints discussed earlier ($\mathcal{C}_5$ in Section \ref{ssec:UAV energy}), creates a challenging solution space to navigate.

\par The interdependencies between the optimization variables further complicate matters. The optimal UAV position depends on both user association and phase shift configuration, while the optimal phase shifts depend on the relative positions of users and UAVs. This creates a circular dependency that cannot be easily resolved through conventional optimization methods. Consequently, the USBJOP is NP-hard \cite{Burer2012}, making it computationally infeasible to find globally optimal solutions for realistic problem sizes using direct approaches.

\par In the following section, we propose an effective solution methodology that decomposes this complex problem into more manageable subproblems while capturing the essential interdependencies between the optimization variables. This approach enables us to achieve near-optimal performance with reasonable computational complexity, making it suitable for practical UAV-SIM deployments.

\section{Proposed Solution}
\label{sec4}

\par
\begin{figure*}[h] 
    \centering 
    \includegraphics[width=\textwidth]{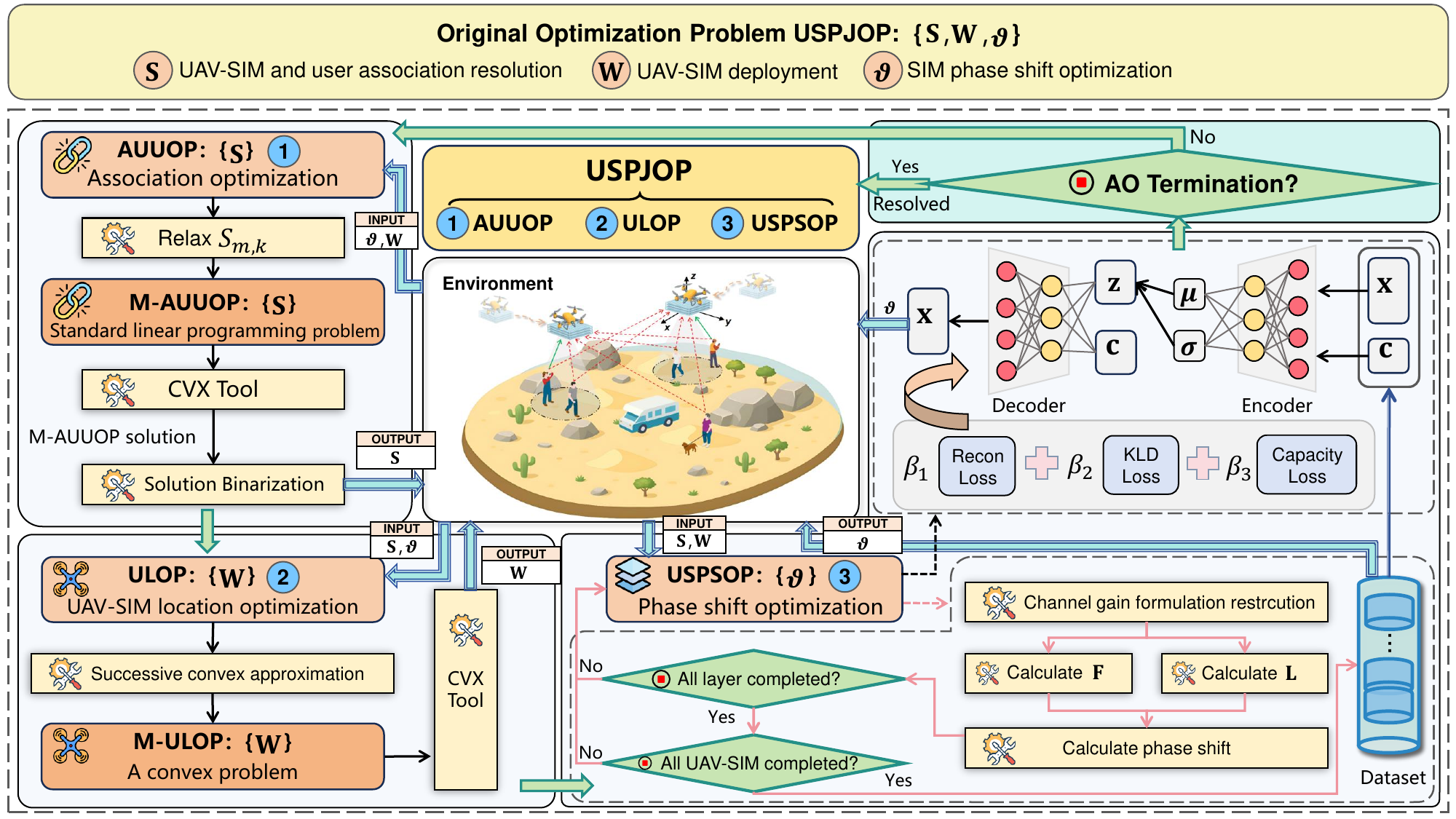} 
    \caption{The structure of the AO strategy framework in LAE networks.} 
    \label{AlgorithmModel} 
    \vspace{-15pt}
\end{figure*}

 In this section, we develop an efficient solution for the formulated USBJOP. Given the inherent complexity of the problem, we employ a divide-and-conquer approach by decomposing USBJOP into three manageable subproblems, which are jointly solved through an AO framework. Specifically, three manageable subproblems, \textit{i.e.}, AUUOP, ULOP, and USPSOP, are formulated to determine the optimal association between UAV-SIMs and ground users, optimize the 3D positioning of UAV-SIMs in the deployment area of LAE networks, and configure the optimal phase shifts across all metasurface layers, respectively. Fig. \ref{AlgorithmModel} provides a visual illustration of this optimization framework, highlighting the interrelationships between the three subproblems and their respective solution approaches, and the details are shown in Algorithm \ref{alg:2}.

\begin{algorithm}[t]
\SetAlgoLined
\SetAlgoNlRelativeSize{-1} 
Let $\tau=0$ and initialize $\boldsymbol{\vartheta}$ and $\mathbf{W}$;\\
\Repeat{$R^{\tau}-R^{\tau-1}\leq\epsilon$ or $\tau=\tau_{\max}$}{
    Update $\mathbf{S}$ to solve AUUOP using CVX tool;\\
    Update $\mathbf{W}$ to solve ULOP using CVX tool;\\
    Update $\boldsymbol{\vartheta}$ to solve USPSOP using HGPSO algorithm;\\
    $\tau=\tau+1$;\\
}
\caption{AO strategy for solving USBJOP\label{alg:2}}
\end{algorithm}

\subsection{AUUOP} 

\subsubsection{AUUOP Formulation}
\label{alpha optimization}

\par Given the UAV-SIM phase shifts and the locations of the UAV-SIMs $\{\boldsymbol{\vartheta},\mathbf{W}\}$, the sub-optimization problem AUUOP is shown as follows: 
\begin{subequations}
    \begin{align}
        \text {(AUUOP)}:\underset{\mathrm{\mathbf{S}}}{\max}\quad & R\\
        \text { s.t. } \quad & \mathcal{C}_2-\mathcal{C}_4.
    \end{align}
\end{subequations}

\par For enhanced tractability of AUUOP, $\mathcal{C}_4$ is relaxed and the binary variable $S_{m, k}$ is relaxed to a continuous one, which satisfies $0 \leq S_{m,k} \leq 1, \forall m\in \mathcal{M}, \forall k\in \mathcal{K}$. The resulting modified AUUOP (M-AUUOP) is given by
\begin{subequations}
    \begin{align}
        \underset{\mathrm{\mathbf{S}}}{\max}\quad & R\\
        \text { s.t. } \quad
        & 0 \leq S_{m,k} \leq 1, \forall m\in \mathcal{M}, \forall k\in \mathcal{K},\\
        & \mathcal{C}_2,\mathcal{C}_3.
    \end{align}
\end{subequations}

\subsubsection{AUUOP Solution}

\par M-AUUOP is a standard linear programming problem. Consequently, it can be readily solved using the CVX tool \cite{Grant2014}. Upon obtaining the solution of M-AUUOP, the association variables are binarized. Specifically, for the $m$-th UAV-SIM, resources are exclusively allocated to the user with the largest service proportion among its associated users from the continuous solution, while all other users are dissociated. This reconstruction process ensures a feasible binary solution.

\subsection{ULOP}

\subsubsection{ULOP Formulation}

\par Given the UAV-SIM phase shifts and the  association between UAV-SIMs and users $\{\boldsymbol{\vartheta},\mathbf{S}\}$, ULOP can be written as follows:
\begin{subequations}
    \label{al2}
    \begin{align}    \text{(ULOP)}:\underset{\mathrm{\mathbf{W}}}{\max}\quad & R\\
        \text { s.t. } \quad
        & \mathcal{C}_1,\mathcal{C}_5. \label{20b}
    \end{align}
\end{subequations}

\subsubsection{ULOP Solution}

\par The non-convexity of constraint $\mathcal{C}_1$ makes ULOP challenging to solve. Successive Convex Approximation (SCA) is then employed to tackle this non-convex constraint \cite{Wu2018}.

\par First, defining $\mathbf{\tilde h}_{m,k}\sim \mathscr{CN}\left(\mathbf{0},\mathbf{R}\right),\forall m\in \mathcal{M}, \forall k\in \mathcal{K}$, the corresponding channel $\mathbf{h}_{m,k}$ can then be rewritten as:
\begin{equation}
    \mathbf{h}_{m, k}=\sqrt{\beta_{m, k}} \tilde{\mathbf{h}}_{m, k}.
\end{equation}
Then, the sub-optimization problem ULOP can be reconstructed as follows:
\begin{equation}
    \begin{split}
        \label{R}
        R= & \sum\nolimits_{m=1}^M \sum\nolimits_{k=1}^K  S_{m,k} \Bigg(\hat{R}_{m,k}-\\
        &\log _2\bigg(\sum\nolimits_{k^{\prime} \neq k}^K \frac{ \rho_0 p_{k^{\prime}}}{\left\|\boldsymbol{w}_m-\boldsymbol{u}_{k^{\prime}}\right\|^2}|\mathbf{w}_m^{1\mathrm{H}} \mathbf{G}_{m}^{\mathrm{H}} \mathbf{\tilde h}_{m,k^{\prime}}|^2+\sigma_m^2\bigg) \Bigg),
    \end{split}
\end{equation}
where $\hat{R}_{m,k} = \log_2\left( \sum\nolimits_{k=1}^K \frac{ \rho_0 p_k \left| \mathbf{w}_m^{1\mathrm{H}} \mathbf{G}_{m}^{\mathrm{H}} \mathbf{\tilde h}_{m,k} \right|^2 }{ \left\| \boldsymbol{w}_m - \boldsymbol{u}_k \right\|^2 } + \sigma_m^2 \right)$.
\par Given the non-convexity of Eq. (\ref{R}) with respect to $\boldsymbol{w}_m$, we let $\mathbf{A}=\{\alpha_{m,k^{\prime}}=\left\|\boldsymbol{w}_m-\boldsymbol{u}_{k^{\prime}}\right\|^2,  k^{\prime}\neq k,\forall k^{\prime} \in \mathcal{K},\forall m \in \mathcal{M}\}$ to relax the problem~\cite{He2025}. Based on this consideration, Eq. (\ref{R}) can be reconstructed as follows:
\begin{subequations}\label{21}
    \begin{align}		\underset{\mathrm{\mathbf{W},\mathbf{A}}}{\max}\quad& R= \sum\nolimits_{m=1}^M \sum\nolimits_{k=1}^K  S_{m,k}\Bigg(\hat{R}_{m,k}-\\
        &\phantom{R=}\log _2\bigg(\sum\nolimits_{k^{\prime} \neq k}^K \frac{\rho_0 p_{k^{\prime}}}{\alpha_{m,k^{\prime}}}|\mathbf{w}_m^{1\mathrm{H}} \mathbf{G}_{m}^{\mathrm{H}} \mathbf{\tilde h}_{m,k^{\prime}}|^2+\sigma_m^2\bigg)\Bigg),\notag\\
        \text { s.t. } \quad
        & \mathcal{C}_1 \label{24c},\mathcal{C}_5\\
        & \mathcal{C}_7:0\leq \alpha_{m,k^{\prime}}\leq\left\|\boldsymbol{w}_m-\boldsymbol{u}_{k^{\prime}}\right\|^2.\label{24b}
    \end{align}
\end{subequations}

\par Given a local point  $\mathbf{W}^\tau= \left\{\boldsymbol{w}_m^\tau, \forall m\in \mathcal{M}, \right\}$  at the $\tau$-th iteration, since any convex function is globally lower-bounded by its first-order Taylor expansion at any point, we can obtain the lower bound of $\hat{R}_{m,k}$ via the first-order Taylor expansion, \textit{i.e.},

\begin{equation}
    \sum_{k=1}^K-A_{m,k}^\tau\left(\left\|\boldsymbol{w}_m-
    \boldsymbol{u}_k\right\|^2-\left\|\boldsymbol{w}_m^\tau-\boldsymbol{u}_k\right\|^2\right)+B_{m,k}^\tau \triangleq \hat{R}^{\prime}_{m,k},
\end{equation}

\noindent where $\hat{R}^{\prime}_{m,k}\leq\hat{R}_{m,k}$. Moreover, $A_{m,k}^\tau$ and $B_{m,k}^\tau$ are the coefficients obtained from the first-order Taylor expansion, corresponding to the linear and constant terms, respectively, which can be written as follows:
\begin{align}
    &A_{m,k}^\tau=\frac{\frac{\rho_0 p_k}{\left(\left\|\boldsymbol{w}_m^\tau-\boldsymbol{u}_k\right\|^2\right)^2}\left|\mathbf{w}_m^{1\mathrm{H}} \mathbf{G}_{m}^{\mathrm{H}} \mathbf{\tilde h}_{m,k}\right|^2 \log _2(\mathrm{e})}{\sum_{l=1}^K \frac{ \rho_0 p_l}{\left\|\boldsymbol{w}_m^\tau-\boldsymbol{u}_l\right\|^2}\left|\mathbf{w}_m^{1\mathrm{H}} \mathbf{G}_{m}^{\mathrm{H}} \mathbf{\tilde h}_{m,l}\right|^2+\sigma_{m}^2},\\
    &B_{m,k}^\tau=\log _2\left(\sum\nolimits_{l=1}^K \frac{\rho_0 p_l}{\left\|\boldsymbol{w}_m^\tau-\boldsymbol{u}_l\right\|^2}\left|\mathbf{w}_m^{1\mathrm{H}} \mathbf{G}_{m}^{\mathrm{H}} \mathbf{\tilde h}_{m,l}\right|^2+\sigma_{m}^2\right).
\end{align}

\par Similarly, we can obtain a tighter constraint for $\mathcal{C}_1$ and $\mathcal{C}_7$ at the given $\mathbf{W^\tau}$ as
\begin{align}
    &\mathcal{C}_8:D_{\min }^2 \leq-\left\|\boldsymbol{w}_i^\tau-\boldsymbol{w}_j^\tau\right\|^2 +2\left(\boldsymbol{w}_i^\tau-\boldsymbol{w}_j^\tau\right)^{\mathrm{T}}\left(\boldsymbol{w}_i-\boldsymbol{w}_j\right),\\
    \mathcal{C}&_9:0\leq \alpha_{m,k^{\prime}} \leq\left\|\boldsymbol{w}_m^\tau-\boldsymbol{u}_{k^{\prime}}\right\|^2+2\left(\boldsymbol{w}_m^\tau-\boldsymbol{u}_{k^{\prime}}\right)^{\mathrm{T}}\left(\boldsymbol{w}_m-\boldsymbol{w}_m^\tau\right).
\end{align}

\par Since $D_{\text{distance}}$ in $\mathcal{C}_5$ can be represented as $\left\|\boldsymbol{w}_m-\boldsymbol{w}_m^0\right\|$, where $\boldsymbol{w}_m^0$ represents the initial location of the $m$-th UAV-SIM. Hence, $\mathcal{C}_5$ can be rewritten as:
\begin{equation}
\mathcal{C}_{10}: 
\frac{P^{\text{prop}}_{m}}{v_m}\left\|\boldsymbol{w}_m-\boldsymbol{w}_m^0\right\| \leq E_{\text{battery}} - \left( P_m^{\text{hover}} + P_m^{\text{extra}} \right) T_{\text{operation}},
\end{equation}
which is clearly a convex constraint. This energy constraint $\mathcal{C}_{10}$ is enforced as a hard constraint within each iteration of our alternating optimization procedure, which ensures that every UAV position update remains within the energy budget and makes the solutions immediately applicable to real-world deployments.

\par Therefore, the modified ULOP (M-ULOP) is approximately formulated as the following problem:
\begin{subequations}\label{27}
    \begin{align}
        \underset{\mathrm{\mathbf{W},\mathbf{A}}}{\max}\quad &R= \sum\nolimits_{m=1}^M \sum\nolimits_{k=1}^K  S_{m,k}\Bigg(\hat{R}^{\prime}_{m,k}- \\ 
        &\phantom{R=}\log _2\bigg(\sum\nolimits_{k^{\prime} \neq k}^K \frac{\rho_0 p_{k^{\prime}}}{\alpha_{m,k^{\prime}}}|\mathbf{w}_m^{1\mathrm{H}} \mathbf{G}_{m}^{\mathrm{H}} \mathbf{\tilde h}_{m,k^{\prime}}|^2+\sigma_m^2\bigg)\Bigg)\notag\\
        \text { s.t. } \quad
        & \mathcal{C}_8-\mathcal{C}_{10}.
    \end{align}
\end{subequations}

\par With this transformation, as shown in~\cite{Wu2018}, M-ULOP is jointly concave with respect to $\boldsymbol{w}_m$ and $\alpha_{m,k^{\prime}}$. Furthermore, constraint $\mathcal{C}_8$ and $\mathcal{C}_9$ are linear, while $\mathcal{C}_{10}$ is a second-order cone programming constraint, which is convex. As a result, M-ULOP is formulated as a convex optimization problem. Similar to AUUOP, it can be efficiently solved using CVX tool. The solution obtained from M-ULOP then serves as an approximation for the original ULOP problem.

\subsection{USPSOP}\label{phase optimization algorithm}

\subsubsection{USPSOP Formulation}
Given the association between UAV-SIMs and users and the locations of the UAV-SIMs $\{\mathbf{S},\mathbf{W}\}$, the sub-optimization problem USPSOP can be written as follows:
\begin{subequations}
    \begin{align}
        \text { (USPSOP) }:\underset{\mathrm{\boldsymbol{\vartheta}}}{\max}\quad & R\\
        \text { s.t. } \quad
        & \mathcal{C}_6. 
    \end{align}
\end{subequations}

\vspace{7pt}

\subsubsection{USPSOP Solution}
\par The core challenge of USPSOP fundamentally lies in the complex joint optimization of $L \times N$ atomic phase shifts, which presents significantly greater complexity than conventional single-layer RIS optimization. Specifically, the operational mechanism of SIM itself introduces substantial mathematical complexity through its cascaded modulation architecture, where the output of each layer serves as the input to the subsequent layer, creating strong interdependencies among phase variables that prevent the direct application of traditional RIS optimization methods. This architectural difference results in exponential growth in computational complexity, while single-layer RIS optimization involves only $N$ independent phase variables, multi-layer SIM requires the simultaneous optimization of $L \times N$ highly coupled variables. Moreover, the resulting optimization problem is inherently non-convex, making direct analytical solutions intractable. These theoretical challenges are further compounded by practical constraints, as UAV-SIMs and ground users in LAE communication networks typically operate under severe computational limitations. Consequently, developing computationally efficient algorithms that can achieve superior phase shift optimization performance within these resource constraints becomes the key to solving USPSOP effectively.
\par To effectively address the challenges posed by USPSOP, we propose an innovative hybrid generative phase shift optimization (HGPSO) algorithm. HGPSO can intelligently select and switch between two distinct sub-algorithms based on different communication scenarios and system configurations. This approach aims to achieve an optimal balance between optimization performance and computational efficiency. The two core sub-algorithms are the layer-by-layer iterative phase shift optimization (LBL-IPSO) algorithm~\cite{Li2024} and a modified powerful GAI model~\cite{Wang2025a}, specifically a conditional variational autoencoder (CVAE)~\cite{Li2025a}, namely CVAE-M. By combining these two methods, HGPSO enhances exploration capability and accelerates convergence while maintaining solution quality, making it more suitable for addressing phase configuration problems of SIM in dynamic environments. In the following sections, we will detail these two core sub-algorithms of HGPSO.

\subsubsection{LBL-IPSO}
\par For the LBL-IPSO algorithm, the core idea is to obtain a suboptimal solution through a phase alignment optimization strategy. Specifically, we maximize the channel gain of numerator, denoted as $\mathbf{w}_m^{1\mathrm{H}} \mathbf{G}_{m}^{\mathrm{H}} \mathbf{h}_{m,k},\forall m \in  \mathcal{M},\forall k \in  \mathcal{K}$, by disregarding the denominator of SINR.
\par Specially, for the $m$-th UAV-SIM, we optimize the phase shift $\boldsymbol{\Phi}^l_{m}$ of the $l$-th layer while keeping the phases of the other $(L-1)$ layers fixed. The channel gain $\mathbf{w}_m^{1\mathrm{H}} \mathbf{G}_{m}^{\mathrm{H}} \mathbf{h}_{m,k}$ can be expanded as follows:
\begin{equation}
    \mathbf{w}_m^{1\mathrm{H}} \mathbf{G}_{m}^{\mathrm{H}} \mathbf{h}_{m,k}=\mathbf{w}_m^{1\mathrm{H}}\boldsymbol{\Phi}^{1\mathrm{H}}_{m}\mathbf{W}^{2\mathrm{H}}_{m}\boldsymbol{\Phi}^{2\mathrm{H}}_{m}\ldots\mathbf{W}^{l\mathrm{H}}_{m}\boldsymbol{\Phi}^{2\mathrm{H}}_{m}\mathbf{h}_{m,k},
\end{equation}
\par This can be equivalently rewritten as
\begin{equation}
    \mathbf{w}_m^{1\mathrm{H}} \mathbf{G}_{m}^{\mathrm{H}} \mathbf{h}_{m,k}=\mathbf{F}^{l\mathrm{H}}_m\boldsymbol{\Phi}^{l\mathrm{H}}_{m}\mathbf{L}^{l\mathrm{H}}_m,
\end{equation}
where $\mathbf{F}^{l\mathrm{H}}_m$ and $\mathbf{L}^{l\mathrm{H}}_m$ can be expressed as
\begin{align}
    &\mathbf{F}^{l\mathrm{H}}_m=\mathbf{w}_m^{1\mathrm{H}}\boldsymbol{\Phi}^{1\mathrm{H}}_{m}\mathbf{W}^{2\mathrm{H}}_{m}\boldsymbol{\Phi}^{2\mathrm{H}}_{m}\ldots\mathbf{W}^{l\mathrm{H}}_{m},\label{31}\\
    \mathbf{L}^{l\mathrm{H}}_m&=\mathbf{W}_m^{(l+1)\mathrm{H}} \boldsymbol{\Phi}_m^{(l+1)\mathrm{H}} \ldots \mathbf{W}_m^{L\mathrm{H}} \boldsymbol{\Phi}_m^{L\mathrm{H}} \mathbf{h}_{m,k}\label{32}.
\end{align}
\par Consequently, the phase can be optimized as follows:
\begin{equation}
    \boldsymbol{\Phi}_m^{l}=\operatorname{diag}\left\{\mathrm{e}^{-j\left(\angle \mathbf{F}^{{l}}_m-\angle \mathbf{L}^{l\mathrm{H}}_m\right)}\right\}\label{33}.
\end{equation}
\par Similarly, the phase of each layer is optimized sequentially using this approach. To ensure effective optimization, we perform $\kappa_{\max}$ iterations for the phase optimization of each layer. The detail of the LBL-IPSO algorithm is provided in Algorithm 2.

\begin{algorithm}[h]
\DontPrintSemicolon
\SetAlgoNlRelativeSize{-1} 
Let $\kappa=1$ and initialize  $\boldsymbol{\Phi}_m^1$,  $\boldsymbol{\Phi}_m^2$, ...,  $\boldsymbol{\Phi}_m^L$ randomly;\\
\For{$m=1$ to $M$}{
    \For{$\kappa=1$ to $\kappa_{\max}$}{
        \For{$l=1$ to $L$}{
            Calculate $\mathbf{F}^{l\mathrm{H}}_m$ according to Eq. (\ref{31});\\	
            Calculate $\mathbf{L}^{l\mathrm{H}}_m$ according to Eq. (\ref{32});\\
            Calculate $	\boldsymbol{\Phi}_m^{l}$ according to Eq. (\ref{33});\\
        }
    }
}
\caption{The layer-by-layer iterative phase shift optimization method \label{algo:3}}
\end{algorithm}
\vspace{-6pt}

\subsubsection{CVAE-M}

\par Since LAE application scenarios grow in complexity, LBL-IPSO faces significant computational bottlenecks that limit its practical applicability. The time complexity of the algorithm scales as $O(M \cdot L \cdot N \cdot \kappa_{\max})$, with triple-nested loops becoming prohibitively expensive as the number of UAV-SIMs ($M$), SIM layers ($L$), or required iterations ($\kappa_{\max}$) increases. This computational burden creates an urgent need for alternative approaches that can deliver comparable optimization performance with substantially reduced computation time.

\par As such, we also propose CVAE-M, a modified conditional variational autoencoder for phase shift optimization in complex UAV-SIM deployments within LAE networks. The motivation for adopting a CVAE-based approach is its ability to learn the underlying mapping between system conditions and optimal phase configurations, enabling direct generation of high-quality solutions without expensive iterative calculations. Once trained, CVAE-M can produce optimized phase shifts in a single forward pass, reducing computation time for dynamic UAV networks.

\par Specifically, CVAE-M builds upon the traditional CVAE framework, which consists of an encoder network that maps input data $\mathbf{x}$ (representing phase configurations) and condition variable $\mathbf{c}$ (containing UAV-SIM positions, user locations, and channel conditions) to a latent distribution, which is given by
\begin{equation}
    q_{\phi}(\mathbf{z}|\mathbf{x},\mathbf{c}) = \mathcal{N}(\mathbf{z}|\boldsymbol{\mu}_{\phi}(\mathbf{x},\mathbf{c}), \boldsymbol{\sigma}^2_{\phi}(\mathbf{x},\mathbf{c})).
\end{equation}

\par While traditional CVAEs optimize for reconstruction accuracy and latent space regularization, they lack direct alignment with our objective of maximizing network capacity. This misalignment represents a fundamental limitation when applying standard deep generative models to wireless optimization problems. To overcome this limitation, we introduce a novel capacity-aware loss function that explicitly incorporates network performance metrics, \textit{i.e.},
\begin{equation}
    \mathcal{L}_{\text{CVAE-M}} = \beta_1 \mathcal{L}_{\text{recon}} + \beta_2 \mathcal{L}_{\text{KL}} + \beta_3 \mathcal{L}_{\text{capacity}}.
    \label{eqLoss}
\end{equation}

\par As can be seen, the critical innovation lies in $\mathcal{L}_{\text{capacity}}$, which measures the squared difference between the network capacity achieved by the generated phase shifts of CVAE-M and that of LBL-IPSO solutions, which is given by
\begin{equation}
    \mathcal{L}_{\text{capacity}} = \frac{1}{N} \sum_{i=1}^{N} (C_{\text{CVAE-M},i} - C_{\text{LBL-IPSO},i})^2.
\end{equation}

\par This capacity-aware loss function transforms CVAE-M from a pure generative model into an optimization-oriented framework that specifically targets network performance metrics. The weighting factors $\beta_1$, $\beta_2$, and $\beta_3$ balance these competing objectives, ensuring the model simultaneously maintains generative properties while optimizing for communication performance.

\par By using solutions from LBL-IPSO as training data, CVAE-M effectively distills the optimization intelligence from the analytical method into a learning-based framework. This knowledge transfer creates a powerful synergy, \textit{i.e.}, LBL-IPSO provides optimization expertise, while CVAE-M offers computational efficiency.

\subsection{Computational Complexity Analysis}

\par For AUUOP, it contains at most $MK$ variables and $M+K+2MK$ constraints. According to \cite{Wang2022}, the number of iterations is $\sqrt{MK} \log _2({1}/{\varepsilon_1})$, where $\varepsilon_1$ is the precision of solving by CVX. Thus, the computational complexity for AUUOP is calculated as $\mathcal{O}[(MK)^{2.5}\log _2(1/\varepsilon_1)]$.

\par Similarly, for ULOP, the total number of variables for $\mathrm{\mathbf{W}}$ and $\mathrm{\mathbf{A}}$ is $MK+2M$. The number of constraints in ULOP is $M(K-1)+M(M-1)/2$, and the number of iterations is $\sqrt{MK+2M} \log _2\big({1}/{\varepsilon_2}\big)$, where $\varepsilon_2$ denotes the precision of CVX when solving ULOP. Thus, the computational complexity for ULOP is calculated as $\mathcal{O}[(MK)^{1.5}(MK+M^2)\log _2{(1/\varepsilon_2)}]$.

\par For USPSOP, when the HGPSO algorithm uses the LBL-IPSO strategy, calculating the $\mathbf{F}^{l\mathrm{H}}_m$ in line $5$ of Algorithm \ref{algo:3} requires $4(l-1)N^2$ floating-point computations, and calculating the $\mathbf{L}^{l\mathrm{H}}_m$ in line $6$ of Algorithm \ref{algo:3} requires $4(L-l)N^2$ floating-point computations. Therefore, the total computational complexity of LBL-IPSO is $\mathcal{O}(4N^2ML^2\kappa_{max})$. Consequently, the total computational complexity for Algorithm \ref{alg:2} is $\mathcal{O}\{[(MK)^{2.5}\log _2(1/\varepsilon_1)+(MK)^{1.5}(MK+M^2)\log _2(1/\varepsilon_2)+4N^2ML^2\kappa_{max}]\tau_{\max}\}$.

\par When using the CVAE strategy, the training complexity of CVAE depends on the dataset size $D_n$, the dimension of the latent variable $z_n$, and the number of neural network layers $L_{n}$. Specifically, training CVAE involves backpropagation, which thereby requires complexity $\mathcal{O}(D_nz_nL_n)$. In the generation phase, the required complexity is $\mathcal{O}(NL)$, and thus the total computational complexity is $\mathcal{O}\{[(MK)^{2.5}\log _2(1/\varepsilon_1)+(MK)^{1.5}(MK+M^2)\log _2(1/\varepsilon_2)+NL]\tau_{\max}\}$.

\par \par Table \ref{ComplexityTable} summarizes the computational complexity comparison of different approaches discussed in our work. It is evident that HGPSO using CVAE has lower computational complexity.
\begin{table}[h]
\centering
\caption{Computational complexity comparison of different algorithms.}
\label{tab:3}
\renewcommand{\arraystretch}{1}
\begin{tabular}{|Sl|Sl|}
\hline
\parbox[c]{5cm}{\centering \textbf{Algorithm}} & 
\parbox[c]{5cm}{\centering \textbf{Time Complexity}} \\
\hline
\parbox[c]{5cm}{\centering Proposed HGPSO using LBL-IPSO} & 
\parbox[c]{5cm}{\centering $\mathcal{O}(4N^2ML^2\kappa_{\max})$} \\
\hline
\parbox[c]{5cm}{\centering Proposed HGPSO using CVAE} & 
\parbox[c]{5cm}{\centering $\mathcal{O}(D_n z_n L_n)$} \\
\hline
\end{tabular}
\label{ComplexityTable}
\end{table}

\section{Simulation Results}
\label{sec5}

\par In this section, we conduct key simulations to demonstrate the effectiveness of the proposed AO framework.

\subsection{Simulation Setups}
\par Regarding parameter configuration, we consider that $K=5$ ground users are distributed randomly in a square area with dimensions of $1000$ m $\times 1000$ m. Moreover, $3$ UAV-SIMs are flying in an aerial region at a fixed height $H=50$ m~\cite{Yang2021}. For the UAV energy consumption model parameters, we adopt representative values for a typical UAV platform as detailed in Table \ref{UAV_Parameter}~\cite{Zeng2019}. The safety distance is established as $d_{\mathrm{min}}=100$ m, and the thickness of the SIM is established at $T_{\mathrm{SIM}}=5\lambda$~\cite{An2023}. Therefore, for an $L$-layer SIM, the spacing between layers is $\delta=T_{\mathrm{SIM}}/L$. Additionally, we specify that the meta-atoms of the SIM are arranged in a square configuration, with each meta-atom having dimensions of $d_{x}d_{y}=(\lambda/2)^2$~\cite{An2023a}.
\par We consider the radio frequency of $28$ GHz, and the corresponding wavelength is $\lambda=10.7$ mm~\cite{An2023a}. Furthermore, the number of meta-atoms for each layer of the SIM is established at $N=36$. The noise power is assumed to be $\sigma^2_m=-110\text{ dBm}, \forall m\in \mathcal{M}$, and the transmit power of each user is configured at $500\text{ mW}$. Moreover, the channel gain at a reference distance of 1 m is established at $\rho_0=(\lambda/4\pi)^2$~\cite{An2023}. We establish the threshold $\epsilon$ for the AO strategy at $10^{-6}$, and the maximum number of iterations for AO is configured as $\tau_{\max}=50$. In addition, for Algorithm \ref{algo:3}, we establish the number of iterations as $\kappa_{\max}=200$, thereby ensuring the quality of the solution. For the CVAE-M training, we utilize a dataset consisting of approximately 80,000 distinct entries collected from various simulation scenarios with comprehensive parameter variations including UAV-SIM positions, user positions, and channel environments to ensure dataset diversity and model generalization capability. 

\begin{table}[h]
\centering
\caption{Notations and simulation values for UAV energy consumption model}
\renewcommand{\arraystretch}{1.2}
\setlength{\tabcolsep}{6pt}
\begin{tabular}{|c|l|c|}
\hline
\textbf{Notation} & \textbf{Physical meaning} & \textbf{Simulation value} \\
\hline
$W$       & UAV-SIM weight (N) & 120 \\
$\rho$    & Air density (kg/m$^3$) & 1.225 \\
$R$       & Rotor radius (m) & 0.5 \\
$A$       & Rotor disc area, $A = \pi R^2$ (m$^2$) & 0.79 \\
$\Omega$  & Rotor angular velocity (rad/s) & 400 \\
$U_{tip}$ & Tip speed of the rotor blade, $U_{tip}=\Omega R$ (m/s) & 200 \\
$s$       & Rotor solidity & 0.05 \\
$\delta$  & Profile drag coefficient & 0.012 \\
$d_{0}$   & Equivalent flat plate area ratio & 0.3 \\
$P_{0}$   & Profile power in hover (W) & 580.7 \\
$v_{0}$   & Mean rotor induced velocity in hover (m/s) & 7.87 \\
$v_{m}$   & UAV flying speed (m/s) & 10.0 \\
$P_{i}$   & Induced power in hover (W) & 944.9 \\
\hline
\end{tabular}
\label{UAV_Parameter}
\end{table}

\subsection{Benchmarks}

\par For comparison purposes, we design and introduce several benchmarks, which are detailed as follows:

\begin{itemize}

\item \textit{Random deployment (RD)}:
In RD, the solution of USBJOP is generated randomly. To enhance the quality of the results, we generate $100$ solutions and subsequently select the one with the optimal performance.

\item \textit{Uniform deployment (UD)}:
We divide the entire horizontal area into $U$ sub-regions, with each UAV-SIM located at the center of the corresponding sub-region, while maintaining a fixed height $H$. For the remaining two sub-optimization problems, we apply the corresponding solution in Section \ref{sec3}, which is consistent with our approach. This strategy is frequently implemented in UAV deployment scenarios \cite{Pan2023}.

\item \textit{Evolutionary algorithms}:
We select two classical evolutionary algorithms, which are differential evolution (DE) and particle swarm optimization (PSO) for comparison \cite{Pan2023}, and configure the number of iterations to $50$. 

\item \textit{Without SIM}:
We consider the scenario where each UAV is not equipped with an SIM, thereby serving uplink communications for users. The UAVs are deployed in a uniform manner throughout the coverage area.
\end{itemize}

\subsection{Optimization Results}

\par In this subsection, we conducted various simulations to evaluate the effectiveness of the proposed AO strategy and HGPSO.

\begin{figure*}[t!] 
\centering
\subfigure[]{\includegraphics[width=0.585\textwidth]{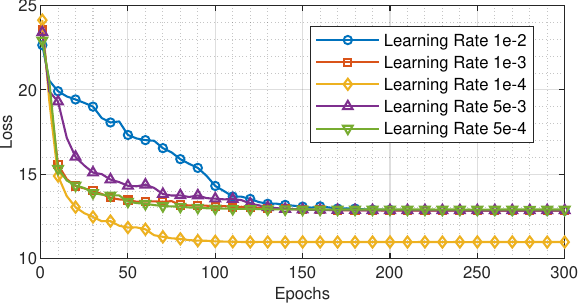}\label{TrainingCompareFigure}}
\subfigure[]{\includegraphics[width=0.39\textwidth]{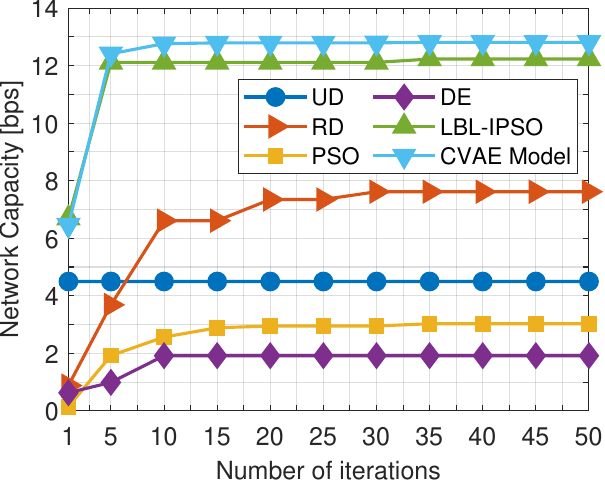}\label{IterationCompareFigure}}
\caption{Algorithm convergence effect diagram. (a) Loss curve diagram of CVAE model under different learning rates. (b) Convergence curves of different algorithms.}
\end{figure*}

\begin{figure}[!t] 
\centering 
\includegraphics[width=3.3 in]{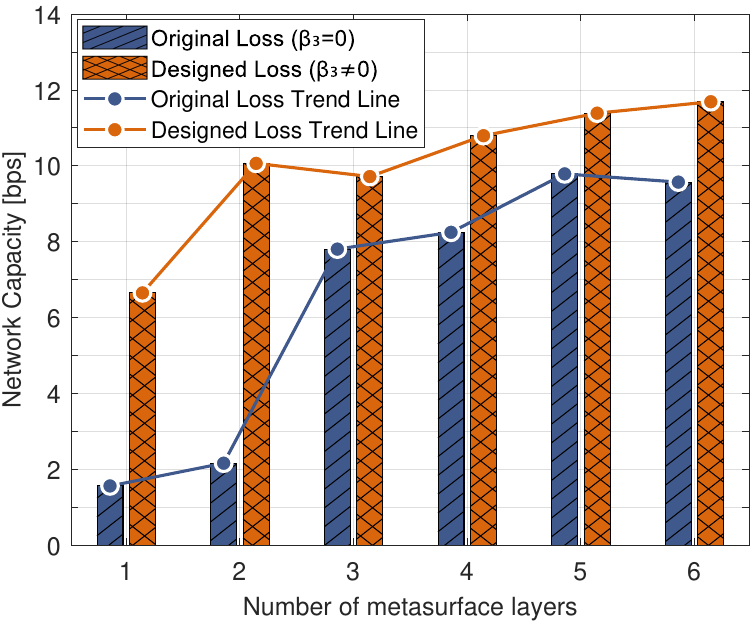} 
\caption{Network performance versus different loss functions.} 
\label{LossCompare} 
\vspace{-5pt}
\end{figure}

\par Fig. \ref{TrainingCompareFigure} illustrates the behavior of the loss function of the CVAE model over $300$ epochs with varying learning rates, where $L=4$ and $N=36$. As can be seen, the results demonstrate that all of the different learning rate configurations start around $23$. In particular, the learning rate of $1e-4$ performs best, with the lowest convergence value of around $11$, while the other four learning rates converge around $13$. Moreover, the $1e-4$ learning rate decreases rapidly in the first $50$ epochs and thereby converges at $100$ epochs. In contrast, the $1e-2$ learning rate shows a slower decline in loss values and requires approximately $150$ epochs to reach relative stability. In addition, learning rates of $1e-3$, $5e-3$ and $5e-4$ exhibit similar convergence patterns. However, the $5e-4$ learning rate shows a faster initial decline rate than the other two. Since the learning rate of $1e-4$ achieved the best results, we will use this learning rate for the CVAE model in subsequent simulations.

\par Fig. \ref{IterationCompareFigure} shows the network capacity of various algorithms after $50$ iterations, where $L=4$ and $N=36$. As can be seen, the CVAE model and LBL-PSO demonstrate outstanding performance, with both algorithms converging rapidly to approximately $12$ bps after just $5$ iterations. In contrast, RD exhibits a noticeable gradual improvement pattern, thereby reaching a maximum value of around $7.5$ by the $30$-th iteration. Since UD uses uniform deployment, its value remains consistently at $4.5$. Both swarm intelligence optimization algorithms, PSO and DE, achieve stable convergence, but their performance thus falls considerably short of the performance of the CVAE Model and LBL-PSO.

\par To further validate the effectiveness of our proposed approach, we conduct ablation studies demonstrating how different components of our loss function in Eq. (\ref{eqLoss}) affect the overall performance. Specifically, we perform comparative simulations between the standard CVAE loss function and our improved formulation. The standard CVAE loss function focuses solely on reconstruction accuracy and latent space regularization, while our designed loss function incorporates an additional capacity-oriented term controlled by parameter $\beta_3$. The comparative results, as shown in Fig. \ref{LossCompare}, demonstrate significant performance advantages of our designed loss function, achieving approximately $400$\% improvement in network capacity for single metasurface layer configurations compared to the standard approach. This performance gap persists across all configurations. These results validate that traditional CVAE require domain-specific modifications to excel in wireless optimization problems.

\begin{figure*}[t!] 
    \centering
    \subfigure[]{\includegraphics[width=0.47\textwidth]{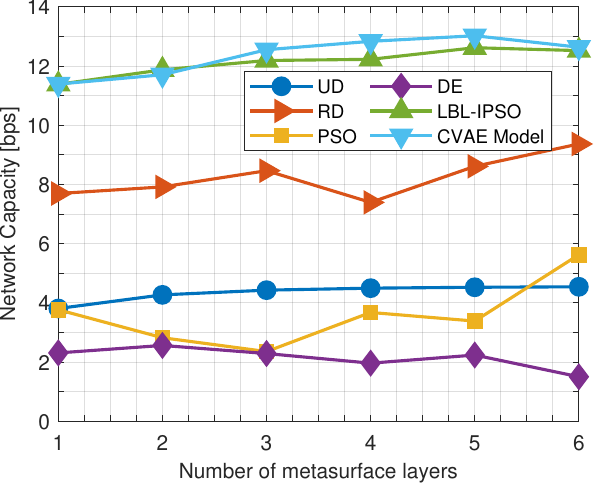}\label{LayerCompareFigure}}
    \subfigure[]{\includegraphics[width=0.47\textwidth]{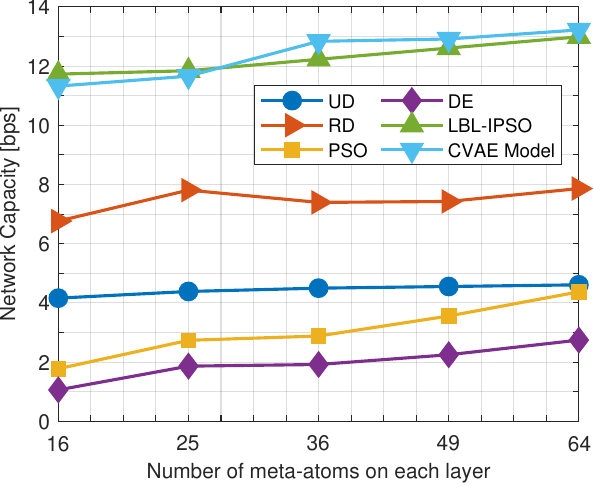}\label{AtomCompareFigure}}
    \subfigure[]{\includegraphics[width=0.47\textwidth]{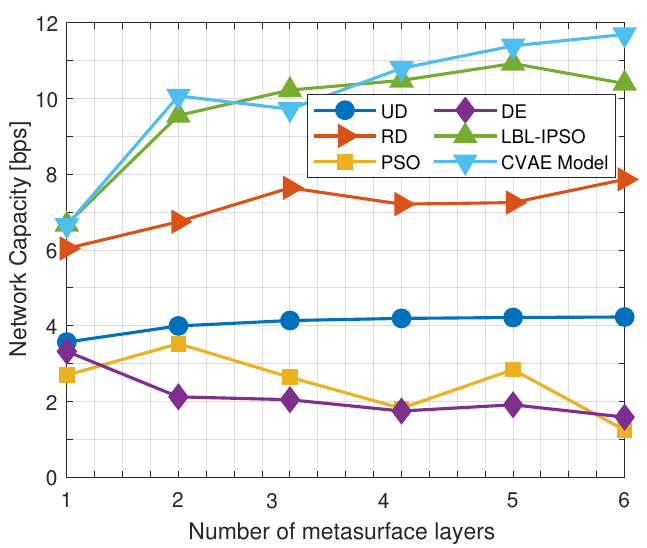}\label{58Layer}}
    \subfigure[]{\includegraphics[width=0.47\textwidth]{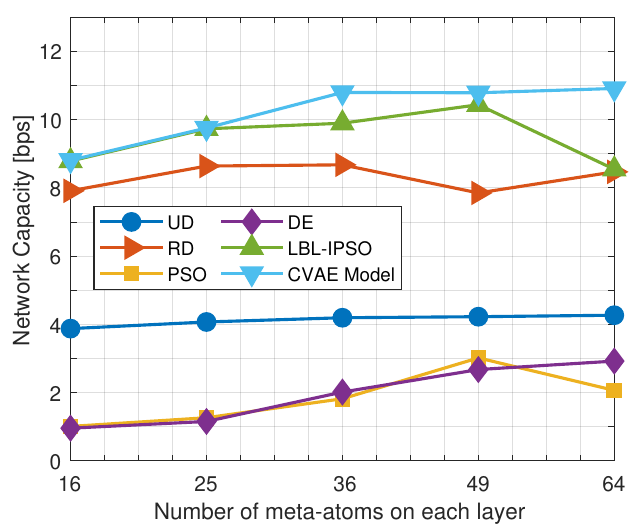}\label{58Atom}}
    \caption{Network capacity $R$ versus SIM configuration. (a) Network capacity $R$ versus the number of metasurface layers $L$ with $K=5$ and $M=3$. (b) Network capacity $R$ versus the number of meta-atoms on each layer $N$ with $K=5$ and $M=3$. (c) Network capacity $R$ versus the number of metasurface layers $L$ with $K=8$ and $M=5$. (d) Network capacity $R$ versus the number of meta-atoms on each layer $N$ with $K=8$ and $M=5$. }
    \label{fig:three_side_by_side}
\vspace{-5pt}
\end{figure*}

\par \par Fig. \ref{LayerCompareFigure} shows a comparison of network capacity performance across various algorithms at different SIM layer counts, where $N = 36$. The CVAE and LBL-IPSO algorithms perform best across all layer counts. Between $1$ and $5$ layers, both methods exhibit a positive correlation with increasing SIM layer count, thereby reaching approximately $13$ bps at $L=5$. The performance declines slightly when the layer count becomes too large. Among these, CVAE performs slightly better than LBL-IPSO in layers $3$ to $5$. Although the RD algorithm performs well, it still lags behind CVAE and LBL-IPSO and exhibits significant variability, which is attributed to the random nature of RD. In the UD algorithm, performance increases from $3.8$ bps to $4.5$ bps as the number of layers increases, thus further indicating that increasing the number of SIM layers within a reasonable range can enhance the network capacity of the system. The swarm intelligence optimization algorithms PSO and DE perform relatively poorly among all comparison algorithms, which are significantly less effective than CVAE and LBL-IPSO. This phenomenon may be attributed to the large solution space, which consequently leads to suboptimal algorithm performance. To evaluate scalability, Fig. \ref{58Layer} shows similar trends in an expanded scenario with $K=8$ users and $M=5$ UAV-SIMs, where our CVAE model continues to outperform all benchmark methods and reaches nearly $12$ bps with $6$ layers, representing a $14.3\%$ improvement over LBL-IPSO. This demonstrates the effective scalability of our proposed approach across varying network complexities.

\par Next, we explore the possible impact of different SIM configurations on network capacity from the perspective of the number of meta-atoms on each layer. Fig. \ref{AtomCompareFigure} shows the performance of different algorithms in terms of network capacity as the number of meta-atoms varies, where $L = 4$. The CVAE and LBL-IPSO algorithms demonstrate superior results in all configurations, with the network capacity of both algorithms being approximately $11$ bps when $N=16$. The network capacity increases as $N$ increases, thus reaching a maximum value of $13$ bps when $N=64$. This observation indicates that the multi-user interference cancellation capability of the SIM is positively correlated with the number of meta-atoms on each layer, which can be further verified through the trend of UD. The RD algorithm exhibits moderate performance with minor fluctuations, yet the same trend can still be observed. Furthermore, the network capacity under both PSO and DE can be steadily improved as the number of meta-atoms increases, thereby resulting in an enhancement of approximately $2$ bps for both algorithms. In comparison to Fig. \ref{LayerCompareFigure}, it is more advantageous to prioritize increasing the number of meta-atoms when modifying the configuration of the SIM.  Similarly, Fig. \ref{58Atom} demonstrates the scalability of our approach in the larger network scenario, where the CVAE model achieves approximately $11$ bps with $64$ meta-atoms, which confirms the robust performance across both network scaling and metasurface dimensionality.

\begin{figure}[!t] 
\centering 
\includegraphics[width=3.3 in]{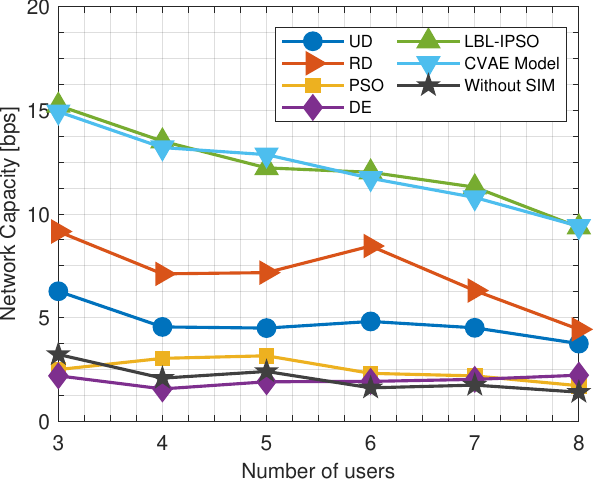} 
\caption{Network capacity $R$ versus the number of ground users $K$.} 
\label{UserCompareFigure} 
\vspace{-5pt}
\end{figure}

\par Fig. \ref{UserCompareFigure} illustrates the network capacity under different numbers of users within LAE networks, where $L=4$ and $N=36$. The network capacity under all the different algorithms decreases as the number of users increases, which is due to the constraint that a UAV-SIM can only serve one user, and a user can only receive one UAV-SIM service. Notably, CVAE and LBL-IPSO continue to be the algorithms with optimal performance, thereby yielding approximately $15$ bps with $3$ users, and the performance gradually decreases to approximately $9.4$ bps as the number of users increases to $8$. It should be noted that the network capacity decreases from $3.2$ bps initially to approximately $1.4$ bps in the absence of SIM service. When $K=3$, the network capacity with SIM is approximately $4.7$ times that without SIM, whereas it increases to approximately $5.4$ times when $K=6$ and to approximately $6.7$ times when $K=8$. This result thus demonstrates that, particularly within multi-user LAE deployment scenarios, the SIM can effectively mitigate the interference caused by multiple users. These substantial performance gains stem from the multi-layer SIM architecture that provides significantly more degrees of freedom for wavefront control and enables superior interference mitigation compared to conventional single-layer designs.

\begin{figure}[!t] 
\centering 
\includegraphics[width=3.2 in]{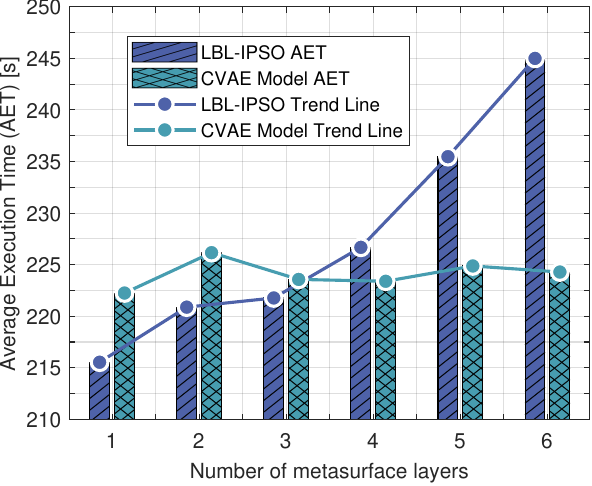} 
\caption{Average execution time comparison of HGPSO with different strategies under AO.} 
\label{TimeCompareFigure} 
\vspace{-5pt}
\end{figure}

\par These results verify the superior performance of CVAE and LBL-IPSO. To further explore the time efficiency of different switching strategies in HGPSO, we compare and analyse their average execution time (AET). Under a unified CPU running environment with $N=36$, $30$ separate runs are performed for different SIM layers to calculate the average value and thus reduce the influence of outliers on the results. Fig. \ref{TimeCompareFigure} shows the trend in AET for the CVAE model and the LBL-IPSO algorithm for different numbers of layers. As the complexity of the LAE scenario increases, it is clear that the two strategies show a significant difference in running efficiency. The AET for the LBL-IPSO algorithm starts at around $215$ seconds for a single layer SIM and increases with the number of layers, thereby reaching approximately $245$ seconds at $6$ layers, which is an increase of about $14\%$. In contrast, the CVAE model consistently maintains an AET between $222$ and $226$ seconds under the same conditions, which demonstrates greater stability with a smaller range of fluctuation. Compared with the LBL-IPSO algorithm, the AET is reduced by approximately $10\%$ at $6$ layers. This difference in trend suggests that the algorithmic efficiency of CVAE can effectively manage the computational burden of complex optimization problems in large-scale LAE network environments without time loss. This is because traditional optimization methods face exponentially growing computational burdens with more SIM layers and meta-atoms, while our CVAE approach shifts the computational burden to an offline training phase and enables rapid online inference. Consequently, this makes CVAE the preferred choice for real-time UAV-SIM deployments in dynamic environments where rapid adaptation is crucial.

\vspace{-10pt}
%
%
\section{Conclusion}
\label{sec6}

\par In this paper, we have investigated a novel communication system architecture that integrates the computational capabilities of multi-layer SIMs with the mobility advantages of UAVs to enhance uplink communications for ground users in LAE networks. We formulated this scenario as  USBJOP that simultaneously optimizes three critical dimensions, which are the association between UAV-SIMs and ground users, the three-dimensional positioning of UAV-SIMs, and the phase shift configurations across multiple SIM layers. To tackle the NP-hardness and non-convexity of the problem, we proposed an AO strategy that decomposes the original problem into three manageable sub-problems, \textit{i.e.}, AUUOP, ULOP, USPSOP. We have solved AUUOP and ULOP using convex optimization techniques via the CVX Tool, while addressing USPSOP with a GAI-based hybrid optimization algorithm. Simulation results demonstrate the superior performance of the proposed architecture and optimization strategy, achieving approximately 1.5 times the network capacity compared to suboptimal benchmarks. The results further reveal that SIMs are highly effective at mitigating multi-user interference, with this effectiveness positively correlated with the number of SIM layers and meta-atoms per layer within specific ranges. Moreover, the proposed GAI method significantly accelerates the solution process in complex scenarios, reducing the algorithm runtime by $10\%$ while maintaining solution quality. 
\vspace{-15pt}
\bibliographystyle{IEEEtran}
\bibliography{reference.bib}
\end{document}